# COMPUTING QUANTUM PHASE TRANSITIONS

Thomas Vojta

Department of Physics, University of Missouri-Rolla, Rolla, MO 65409

## PREAMBLE: MOTIVATION AND HISTORY

A phase transition occurs when the thermodynamic properties of a material display a singularity as a function of the external parameters. Imagine, for instance, taking a piece of ice out of the freezer. Initially, its properties change only slowly with increasing temperature. However, at 0°C, a sudden and dramatic change occurs. The thermal motion of the water molecules becomes so strong that it destroys the crystal structure. The ice melts, and a new phase of water forms, the liquid phase. At the phase transition temperature of 0°C the solid (ice) and the liquid phases of water coexist. A finite amount of heat, the so-called latent heat, is required to transform the ice into liquid water. Phase transitions involving latent heat are called first-order transitions. Another well known example of a phase transition is the ferromagnetic transition of iron. At room temperature, iron is ferromagnetic, i.e., it displays a spontaneous magnetization. With rising temperature, the magnetization decreases continuously due to thermal fluctuations of the spins. At the transition temperature (the so-called Curie point) of 770°C, the magnetization vanishes, and iron is paramagnetic at higher temperatures. In contrast to the previous example, there is no phase coexistence at the transition temperature; the ferromagnetic and paramagnetic phases rather become indistinguishable. Consequently, there is no latent heat. This type of phase transition is called continuous (second-order) transition or critical point.

Phase transitions play an essential role in shaping the world. The large scale structure of the universe is the result of phase transitions during the early stages of its development after the Big Bang. Phase transitions also occur during the production of materials, in growth processes, and in chemical reactions. Understanding phase transitions has thus been a prime endeavor of science. More than a century has gone by from the first (modern) discoveries by Andrews[1] and van der Waals[2] until a consistent picture started to emerge. However, the theoretical concepts established during this development, viz., scaling and the renormalization group[3,4] now belong to the central paradigms of modern physics and chemistry.

The examples of phase transitions mentioned above occur at nonzero temperature. At these so-called thermal or classical transitions, the ordered phase (the ice crystal or the ferromagnetic state of iron) is destroyed by thermal fluctuations. In the last two decades or so, considerable attention has focused on a very different class of phase transitions. These new transitions occur at zero temperature when a non-thermal parameter like pressure, chemical composition or magnetic field is changed. The fluctuations which destroy the ordered phase in these transitions cannot be of thermal nature. Instead, they are quantum fluctuations which are a consequence of Heisenberg's uncertainty principle. For this reason, these zero-temperature transitions are called quantum phase transitions.

As an illustration, the magnetic phase diagram[5] of the compound $LiHoF_4$ is shown in Fig. 1. In zero external magnetic field, $LiHoF_4$ undergoes a phase transition from a paramagnet to a ferromagnet at about 1.5 K. This transition is a thermal continuous phase transition analogous to the Curie point of iron discussed above. Applying a magnetic field perpendicular to the ordering direction of the ferromagnet induces quantum fluctuations between the spin-up and spin-down states and thus reduces the transition temperature. At a field strength of about 50 kOe, the transition temperature drops to zero. Thus, at 50 kOe $LiHoF_4$ undergoes a quantum phase transition from a ferromagnet to a paramagnet. At the first glance, quantum phase transitions seem to be a purely academic problem since they occur at isolated values in parameter space and at zero temperature which is not accessible in a real experiment. However, it has now become clear that the opposite is true. Quantum phase transitions do have important, experimentally relevant consequences, and they are believed to provide keys to many new and exciting properties of condensed matter, including the quantum Hall effects, exotic superconductivity, and non-Fermi liquid behavior in metals.

The purpose of this chapter is twofold: In the following two sections on "*Phase Transitions and Critical Phenomena*" and "*Quantum vs. Classical Phase Transitions*" we give a concise introduction into the theory of quantum phase transitions, emphasizing similarities with and differences to classical thermal transitions. After that, we point out the computational challenges posed by quantum phase transitions, and we discuss a number of successful computational approaches together with prototypical examples. However, this chapter is not meant to be comprehensive in scope. We rather want to help scientists who are doing their first steps in this field to get off on the right foot. Moreover, we want to provide experimentalists and traditional theorists with an idea of what simulations can achieve in this area (and what they cannot do … yet). Those readers who want to learn more details about quantum phase transitions and their applications should consult one of the recent review articles[6,7,8,9] or the excellent text book on quantum phase transitions by Sachdev.[10]

# PHASE TRANSITIONS AND CRITICAL BEHAVIOR

In this section, we briefly collect the basic concepts of the modern theory of phase transitions and critical phenomena to the extent necessary for the purpose of this chapter. A detailed exposition can be found, e.g., in the textbook by Goldenfeld.[11]

**Landau theory**

Most modern theories of phase transitions are based on Landau theory.[12] Landau introduced the concept of an *order parameter*, a thermodynamic quantity that vanishes in one phase (the disordered phase) and is non-zero and generally non-unique in the other phase (the ordered phase). For example, for the ferromagnetic phase transition, the total magnetization is an order parameter. In general, the order parameter can be a scalar, a vector, or even a tensor. Landau theory can be understood as a unification of earlier mean-field theories such as the van-der-Waals theory of the liquid-gas transition[2] or Weiss' molecular field theory of ferromagnetism.[13] It is based on the crucial assumption that the free energy is an analytic function of the order parameter $m$ and can thus be expanded in a power series,

$$F = F_L(m) = F_0 + rm^2 + wm^3 + um^4 + O(m^5) \qquad [1]$$

Close to the phase transition, the coefficients $r,w,u$ vary slowly with the external parameters such as temperature, pressure, electric or magnetic field. For a given system, they can be determined either by a first-principle calculation starting from a microscopic model or phenomenologically by comparison with experimental data. The correct equilibrium value of the order parameter $m$ for each set of external parameter values is found by minimizing $F_L$ with respect to $m$

Let us now discuss the basic properties of phase transitions that result from the Landau free energy [1]. If the coefficient $r$ is sufficiently large, the minimum of $F_L$ is located at $m = 0$, i.e., the system is in the disordered phase. In contrast, for sufficiently small (negative) $r$, the minimum is at some nonzero $m$, putting the system into the ordered phase. Depending on the value of $w$, the Landau free energy [1] describes a first-order or a continuous transition. If $w \ne 0$, the order parameter jumps discontinuously from $m = 0$ to $m \ne 0$, i.e., the transition is of first order. If $w = 0$ (as is often the case due to symmetry), the theory describes a continuous transition or critical point at $r = 0$. In this case, $r$ can be understood as the distance from the critical point, $r \propto T - T_c$. Within Landau theory, the behavior close to a critical point is super-universal, i.e., all continuous phase transitions display the same behavior. For instance, the order parameter vanishes as $m = (-r/2u)^{1/2}$ when the critical point $r = 0$ is approached from the ordered phase, implying that the critical exponent $\beta$ which describes the singularity of the order parameter at the critical point via $m \propto |r|^\beta \propto |T - T_c|^\beta$ always has the mean-field value 1/2.

In experiments, the critical exponent values are in general different from what Landau theory predicts; and while they show some degree of universality, it is weaker than the predicted super-universality. For instance, all three-dimensional Ising ferromagnets (i.e., ferromagnets with uniaxial symmetry and a scalar order parameter) fall into the same universality class with $\beta \approx 0.32$ while all two-dimensional Ising magnets have $\beta \approx 1/8$. All three-dimensional Heisenberg magnets (for which the order parameter is a three-component vector with $O(3)$ symmetry) also have a common value of $\beta \approx 0.35$ but it is different from the one in Ising magnets.

The failure of Landau theory to correctly describe the critical behavior was the central puzzle in phase transition theory over many decades. It was only solved in the 1970's with the development of the renormalization group.[3,4] We now understand that Landau theory does not adequately include the fluctuations of the order parameter about its average value. The effects of these fluctuations in general decrease with increasing dimensionality and with increasing number of order parameter components. This suggests that Landau theory might actually be correct for systems in sufficiently high space dimension $d$. In fact, the fluctuations lead to two different critical dimensionalities, $d_c^+$ and $d_c^-$, for a given phase transition. If $d$ is larger than the upper critical dimension $d_c^+$, fluctuations are unimportant for the critical behavior, and Landau theory gives the correct critical exponents. If $d$ is between the upper and the lower critical dimensions, $d_c^+ > d > d_c^-$, a phase transition still exists but the critical behavior is different from Landau theory. For dimensionalities below the lower critical dimension, fluctuations become so strong that they completely destroy the ordered phase. For the ferromagnetic transition at nonzero temperature, $d_c^+ = 4$, and $d_c^- = 2$ or 1 for Heisenberg and Ising symmetries, respectively.

**Scaling and the Renormalization Group**

To go beyond Landau theory, the order parameter fluctuations need to be included. This can be achieved by writing the partition function as a functional integral

$$Z = e^{-F/k_B T} = \int D[\phi] e^{-S[\phi]/k_B T} \qquad [2]$$

where $S[\phi]$ is the Landau-Ginzburg-Wilson (LGW) free energy functional, a generalization of the Landau free energy [1] for a fluctuating field $\phi(\mathbf{x})$ representing the local order parameter. It is given by

$$S[\phi] = \int d^d x \left[ c \left( \nabla \phi(\mathbf{x}) \right)^2 + F_L(\phi(\mathbf{x})) - h\phi(\mathbf{x}) \right]. \qquad [3]$$

Here, we have also included an external field $h$ conjugate to the order parameter (in the case of the ferromagnetic transition, $h$ is a uniform magnetic field). The thermodynamic average $m$ of the order parameter is given by the average $\langle \phi \rangle$ of the field with respect to the statistical weight $e^{-S[\phi]/k_B T}$.

In the disordered phase, the thermodynamic average of the order parameter vanishes, but its fluctuations are nonzero. When the critical point is approached, the spatial correlations of the order parameter fluctuations, as characterized by the correlation function $G(\mathbf{x} - \mathbf{y}) = \langle \phi(\mathbf{x})\phi(\mathbf{y}) \rangle$, become long-ranged. Close to the critical point, their typical length scale, the correlation length $\xi$, diverges as

$$\xi \propto |r|^{-\nu} \qquad [4]$$

where $\nu$ is called the correlation length critical exponent. This divergence was observed in 1873 in a famous experiment by Andrews[1]: A fluid becomes milky when approaching its critical point because the length scale of its density fluctuations reaches the wavelength of light. This phenomenon is called critical opalescence.

Close to the critical point, $\xi$ is the only relevant length scale in the system. Therefore, the physical properties must be unchanged, if all lengths in the system are rescaled by a common factor *b*, and at the same time the external parameters are adjusted in such a way that the correlation length retains its old value. This gives rise to a homogeneity relation for the free energy density $f = -(k_B T/V)\log Z$,

$$f(r,h) = b^{-d} f(rb^{1/\nu}, hb^{y_h}) \qquad [5]$$

The scale factor *b* is an arbitrary positive number, and $y_h$ is another critical exponent. Analogous homogeneity relations for other thermodynamic quantities can be obtained by taking derivatives of *f*. These homogeneity laws were first obtained phenomenologically[14] and are sometimes summarily called the scaling hypothesis. Within the framework of the modern renormalization group theory of phase transitions[3,4] the scaling laws can be derived from first principles. The diverging correlation length is also responsible for the above-mentioned universality of the critical behavior. Close to the critical point, the system effectively averages over large volumes rendering microscopic system details irrelevant. As a result, the universality classes are determined only by symmetries and the spatial dimensionality.

In addition to the critical exponents $\nu$ and $y_h$ defined above, other exponents describe the dependence of the order parameter and its correlations on the distance from the critical point and on the field conjugate to the order parameter. The definitions of the most commonly used critical exponents are summarized in Table 1. These exponents are not all independent from each other. The four thermodynamic exponents $\alpha, \beta, \gamma, \delta$ all derive from the free energy [5] which contains only two independent exponents. They are therefore connected by the scaling relation

$$\begin{aligned} 2 - \alpha &= 2\beta + \gamma \\ 2 - \alpha &= \beta(\delta + 1) \end{aligned} \qquad [6]$$

The correlation length and correlation function exponents are related by

$$2 - \alpha = d\nu$$
$$\gamma = (2 - \eta)\nu$$
[7]

|  | Exponent | Definition | Conditions |
| --- | --- | --- | --- |
| Specific Heat | $\alpha$ | $c \propto \|r\|^{-\alpha}$ | $r \to 0, h = 0$ |
| Order Parameter | $\beta$ | $m \propto (-r)^{\beta}$ | $r \to 0-, h = 0$ |
| Susceptibility | $\gamma$ | $\chi \propto \|r\|^{-\gamma}$ | $r \to 0, h = 0$ |
| Critical Isotherm | $\delta$ | $h \propto \|m\|^{\delta} \mathrm{sgn}(m)$ | $r = 0, h \to 0$ |
| Correlation Length | $\nu$ | $\xi \propto \|r\|^{-\nu}$ | $r \to 0, h = 0$ |
| Correlation Function | $\eta$ | $G(\mathbf{x}) \propto \|\mathbf{x}\|^{-d+2-\eta}$ | $r = 0, h = 0$ |
| Dynamical | $z$ | $\xi_t \propto \xi^z$ | |
| Activated Dynamical | $\psi$ | $\ln \xi_t \propto \xi^{\psi}$ | |

**Table 1: Definitions of critical exponents.** *m* **is the order parameter, and** *h* **is the conjugate field.** *r* **denotes the distance from the critical point, and** *d* **is the space dimensionality. The exponent** $y_h$ **defined in [5] is related to** $\delta$ **via** $y_h = d\delta/(1+\delta)$.

Exponent relations explicitly involving the dimensionality *d* are called hyperscaling relations. They only hold below the upper critical dimension $d_c^+$. Above $d_c^+$ they are destroyed by dangerously irrelevant variables.[11]

In addition to the diverging length scale $\xi$, a critical point is characterized by a diverging time scale, the correlation time $\xi_t$. It leads to the phenomenon of *critical slowing down*, i.e., very slow relaxation towards equilibrium near a critical point. At generic critical points, the divergence of the correlation time follows a power law $\xi_t \propto \xi^z$ where $z$ is the dynamical critical exponent. At some transitions, in particular in the presence of quenched disorder, the divergence can be exponential, $\ln \xi_t \propto \xi^\psi$. The latter case is referred to as activated dynamical scaling in contrast to the generic power-law dynamical scaling.

**Finite-size scaling**

The question of how a finite system size influences a critical point is very important for the application of computational methods and also for many experiments, e.g., in layered systems or nano materials. In general, a sharp phase transition can only exist in the thermodynamic limit, i.e., in an infinite system. A finite system size results in a rounding and shifting of the critical singularities. A quantitative description of finite size effects is provided by finite-size scaling theory.[15,16,17] Finite-size scaling starts from the observation that the inverse system size acts as an additional parameter (analogous to $r$ or $h$) that takes the system away from the critical point. Because the correlation length of the infinite system $\xi_\infty$ is the only relevant length scale close to the critical point, finite-size effects in a system of linear size $L$ must be controlled by the ratio $L/\xi_\infty$ only. We can therefore generalize the classical homogeneity relation [5] for the free energy density by including the system size

$$f(r,h,L) = b^{-d} f(rb^{1/\nu}, hb^{y_h}, Lb^{-1}) \qquad [8]$$

By taking derivatives and/or setting the arbitrary scale factor $b$ to appropriate values, [8] can be used to derive scaling forms of various observables. For instance, setting $b = L$ and $h = 0$, we obtain $f(r,L) = L^{-d} \Theta_f(rL^{1/\nu})$ where $\Theta_f(x)$ is a dimensionless scaling function. This can also be used to find how the critical point shifts as a function of $L$ in geometries that allow a sharp transition at finite $L$ (e.g., layers of finite thickness). The finite-$L$ phase transition corresponds to a singularity in the scaling function at some nonzero argument $x_c$. The transition thus occurs at $r_c L^{1/\nu} = x_c$, and the transition temperature $T_c(L)$ of the finite-size system is shifted from the bulk value $T_c^0$ by

$$T_c(L) - T_c^0 \propto r_c = x_c L^{-1/\nu} \qquad [9]$$

Note that the simple form of finite-size scaling summarized above is only valid below the upper critical dimension $d_c^+$ of the phase transition. Finite-size scaling can be generalized to dimensions above $d_c^+$, but this requires taking dangerously irrelevant variables into account. One important consequence is that the shift of the critical temperature, $T_c(L) - T_c^0 \propto L^{-\varphi}$ is controlled by an exponent $\varphi$ which in general is different from $1/\nu$.

Finite-size scaling has become one of the most powerful tools of analyzing computer simulation data of phase transitions. Instead of treating finite-size effects as errors to be avoided, one can simulate systems of varying size and test whether homogeneity relations such as [8] are fulfilled. Fits of the simulation data to the finite-size scaling forms of the observables then yield values for the critical exponents. We will discuss examples of this method later in the chapter.

**Quenched disorder**

Realistic systems always contain some amount of quenched (i.e., frozen-in) disorder in the form of vacancies, impurity atoms, dislocations, or other types of imperfections. Understanding their influence on the behavior of phase transitions and critical points is therefore very important for analyzing experiments. In this section, we focus on the simplest type of disorder (sometimes called weak disorder, random-$T_c$ disorder, or, from the analogy to quantum field theory, random-mass disorder) by assuming that the impurities and defects do not qualitatively change the bulk phases that are separated by the transition. They only lead to spatial variations of the coupling strength and thus of the local critical temperature. In ferromagnetic materials, random-$T_c$ disorder can be achieved, e.g., by diluting the lattice, i.e., by replacing magnetic atoms with nonmagnetic ones. Within a LGW theory such as [3], random-$T_c$ disorder can be modeled by making the bare distance from the critical point a random function of spatial position, $r \to r + \delta r(\mathbf{x})$.

The presence of quenched disorder naturally leads to the following questions:
- Will the phase transition remain sharp or will it be rounded?
- Will the order of the transition (1st order or continuous) remain the same as in the clean case?
- If the transition is continuous, will the dirty system show the same critical behavior as the clean one or will the universality class change?
- Will only the transition itself be influenced or also the behavior in its vicinity?

An important early step towards answering these questions is due to Harris[18] who considered the stability of a critical point against disorder. He showed that if a clean critical point fulfills the exponent inequality

$$d\nu > 2,  \qquad [10]$$

now called the Harris criterion, it is perturbatively stable against weak disorder. Note, however, that the Harris criterion only deals with the average behavior of the disorder at large length scales; effects due to qualitatively new behavior at finite length scales (and finite disorder strength) are not covered. Thus, the Harris criterion is a necessary condition for the stability of a clean critical point, not a sufficient one.

The Harris criterion can be used as the basis for a classification of critical points with quenched disorder according to the behavior of the average disorder strength with increasing length scale. Three classes can be distinguished.[19] (i) The first class contains critical points fulfilling the Harris criterion. At these phase transitions, the disorder strength *decreases* under coarse graining, and the system becomes homogeneous at large length scales. Consequently, the critical behavior of the dirty system is identical to that of the clean system. Macroscopic observables are self-averaging at the critical point, i.e., the relative width of their probability distributions vanishes in the thermodynamic limit.[20,21] A prototypical example is the three-dimensional classical Heisenberg model whose clean correlation length exponent is $\nu \approx 0.711$ fulfilling the Harris criterion. (ii) In the second class, the system remains inhomogeneous at all length scales with the relative strength of the disorder approaching a finite value for large length scales. The resulting critical point still displays conventional power-law scaling but it is in a new universality class with exponents that differ from those of the clean system (and fulfill the inequality $d\nu > 2$). Macroscopic observables are *not* self-averaging, but in the thermodynamic limit, the relative width of their probability distributions approaches a size-independent constant. An example in this class is the classical three-dimensional Ising model. Its clean correlation length exponent, $\nu \approx 0.629$, does not fulfill the Harris criterion. Introduction of quenched disorder, e.g., via dilution, thus leads to a new critical point with an exponent of $\nu \approx 0.684$. (iii) At critical points in the third class, the relative magnitude of the disorder counter-intuitively *increases* without limit under coarse graining. At these so-called infinite-randomness critical points, the power-law scaling is replaced by activated (exponential) scaling. The probability distributions of macroscopic variables become very broad (even on a logarithmic scale) with their width diverging with system size. Infinite-randomness critical points have mainly been found in quantum systems, starting with Fisher's seminal work on the random transverse field Ising model[22,23] by means of the Ma-Dasgupta-Hu renormalization group.[24]

The above classification is based on the behavior of the *average* disorder strength at large length scales. However, in recent years it has become clear, that often an important role is played by strong disorder fluctuations and the rare spatial regions that support them. These regions can show local order even if the bulk system is in the disordered phase. Their fluctuations are very slow because they require changing the order parameter in a large volume. Griffiths[25] showed that this leads to a singular free energy not only at the phase transition point but in an entire parameter region around it. At generic thermal (classical) transitions, the contribution of the rare regions to thermodynamic observables is very weak since the singularity in the free energy is only an essential one.[26,27] In contrast, at many quantum phase transitions, rare disorder fluctuations lead to strong power-law quantum Griffiths singularities that can dominate the thermodynamic behavior.[22,23,28,29] In some systems, rare region effects can become so strong that they destroy the sharp phase transition by smearing.[30] A recent review of rare region effects at classical, quantum and nonequilibrium phase transitions can be found in Ref. 31.

# QUANTUM VS. CLASSICAL PHASE TRANSITIONS

In this section, we give a concise introduction into the theory of quantum phase transitions, emphasizing similarities with and differences to classical thermal transitions.

**How Important is Quantum Mechanics?**

The question of how important is quantum mechanics for understanding continuous phase transitions has several facets. On the one hand, one may ask whether quantum mechanics is needed to explain the existence and properties of the bulk phases separated by the transition. This question can only be decided on a case-by-case basis, and very often quantum mechanics is essential as, e.g., for the superconducting phase. On the other hand, one can ask how important quantum mechanics is for the behavior close to the critical point and thus for the determination of the universality class the transition belongs to. It turns out that the latter question has a remarkably clear and simple answer: Quantum mechanics does *not* play any role for the critical behavior if the transition occurs at a finite temperature. It does play a role, however, at zero temperature.

To understand this remarkable result, it is useful to distinguish fluctuations with predominantly thermal and quantum character depending on whether their thermal energy $k_B T$ is larger or smaller than the quantum energy scale $\hbar \omega_c$, where $\omega_c$ is the typical frequency of the fluctuations. As discussed in the last section, the typical time scale $\xi_t$ of the fluctuations generally diverges as a continuous transition is approached. Correspondingly, the typical frequency scale $\omega_c$ goes to zero and with it the typical energy scale

$$\hbar \omega_c \propto |r|^{\nu z} \qquad [11]$$

Quantum fluctuations will be important as long as this typical energy scale is larger than the thermal energy $k_B T$. If the transition occurs at some finite temperature $T_c$, quantum mechanics will therefore become unimportant if the distance $r$ from the critical point is smaller than a crossover distance $r_x$ given by $r_x \propto T_c^{1/\nu z}$. Consequently, we find that the critical behavior asymptotically close to the transition is always classical if the transition temperature $T_c$ is nonzero. This justifies calling all finite-temperature phase transitions classical transitions, even if they occur in an intrinsically quantum-mechanical system. Consider, e.g., the superconducting transition of mercury at 4.2 K. Here, quantum mechanics is obviously important on microscopic scales for establishing the superconducting order parameter, but classical thermal fluctuations dominate on the macroscopic scales that control the critical behavior. In other words, close to criticality the fluctuating clusters become so big (their typical size is the correlation length $\xi$) that they behave classically.

In contrast, if the transition occurs at zero temperature as a function of a non-thermal parameter like pressure or magnetic field, the crossover distance $r_x$ vanishes; and quantum mechanics is important for the critical behavior. Consequently, transitions at zero temperature are called quantum phase transitions. In Fig. 2, we show the resulting schematic phase diagram close to a quantum critical point. As discussed above, sufficiently close to the finite-temperature phase boundary, the critical behavior is purely classical. However, the width of the classical critical region vanishes with vanishing temperature. Thus, an experiment along path (a) at sufficiently low temperatures will mostly observe quantum behavior, with a very narrow region of classical behavior (which may be unobservable) right at the transition. The disordered phase comprises three regions, separated by crossover lines. In the quantum disordered region at low temperatures and $B > B_c$, quantum fluctuations destroy the ordered phase, and the effects of temperature are unimportant. In contrast, in the thermally disordered region, the ordered phase is destroyed by thermal fluctuations while the corresponding ground state shows long-range order. Finally, the so-called quantum critical region is located at $B \approx B_c$ and extends (somewhat counter-intuitively) to comparatively high temperatures. In this regime, the system is critical with respect to *B*, and the critical singularities are cut-off exclusively by the temperature. An experiment along path (b) thus explores the temperature scaling of the quantum critical point. The phase diagram in Fig. 2 applies to systems that have an ordered phase at finite-temperatures. Some systems, such as Heisenberg magnets in two dimensions, display long-range order only at exactly zero temperature. The corresponding schematic phase diagram is shown in Fig. 3. Even though the system is always in the disordered phase at any nonzero temperature, the quantum critical point still controls the crossovers between the three different regions discussed above.

**Quantum Scaling and Quantum-to-Classical Mapping**

In classical statistical mechanics, static and dynamic behaviors decouple: Consider a classical Hamiltonian $H(p_i, q_i) = H_{kin}(p_i) + H_{pot}(q_i)$ consisting of a kinetic part $H_{kin}$ that only depends on the momenta $p_i$ and a potential part $H_{pot}$ that only depends on the coordinates $q_i$. The classical partition function of such a system, $Z = \int dp_i e^{-H_{kin}/k_B T} \int dq_i e^{-H_{pot}/k_B T}$, factorizes in kinetic and potential parts which are independent of each other. The kinetic contribution to the free energy generally does not display any singularities, since it derives from a product of Gaussian integrals. One can therefore study the thermodynamic critical behavior in classical systems using time-independent theories like the Landau-Ginzburg-Wilson theory discussed above. The dynamical critical behavior can be found separately.

In quantum statistical mechanics, the situation is different. The kinetic and potential parts of the Hamiltonian do not commute with each. Consequently, the partition function $Z = \text{Tr}\, e^{-\hat{H}/k_B T}$ does not factorize, and one must solve for the dynamics together with the thermodynamics. The canonical density operator $e^{-\hat{H}/k_B T}$ takes the form of a time-evolution operator in imaginary time, if one identifies $1/k_B T = -it/\hbar$. Thus, quantum mechanical analogs of the LGW theory [3] need to be formulated in terms of space and time dependent fields. A simple example of such a quantum LGW functional takes the form

$$S[\phi] = \int_0^{1/k_B T} d\tau \int d^d x \left[ a\left(\partial_\tau \phi(\mathbf{x},\tau)\right)^2 + c\left(\nabla \phi(\mathbf{x},\tau)\right)^2 + F_L(\phi(\mathbf{x},\tau)) - h\phi(\mathbf{x},\tau) \right] \qquad [12]$$

with $\tau$ being the imaginary time variable. It is describes, e.g., the magnetization fluctuations of an Ising model in a transverse field.

This LGW functional also illustrates another remarkable feature of quantum statistical mechanics. The imaginary time variable $\tau$ effectively acts as an additional coordinate whose extension becomes infinite at zero temperature. A quantum phase transition in *d* space dimensions is thus equivalent to a classical transition in *d*+1 dimensions. This property is called the *quantum-to-classical mapping.* In general, the resulting classical system is anisotropic because space and time coordinates do not enter in the same fashion. A summary of the analogies arising from the quantum-to-classical mapping is given in Table 2.

| Quantum System | Classical System |
|---|---|
| $d$ space, 1 time dimensions | $d+1$ space dimensions |
| coupling constant | classical temperature $T$ |
| inverse physical temperature $1/k_B T$ | Finite size $L_t$ in the "time" direction |
| spatial correlation length $\xi$ | spatial correlation length $\xi$ |
| inverse energy gap $\Delta$ | correlation length $\xi_t$ in "time" direction |

**Table 2:** Quantum-to-classical mapping: Analogies between important quantities (after Ref. 6).

The homogeneity law [5] for the free energy can be easily generalized to the quantum case (see, e.g., Ref. 10). For the generic case of power-law dynamical scaling, it takes the form

$$f(r,h,T) = b^{-(d+z)} f(rb^{1/\nu}, hb^{y_h}, Tb^z) \qquad [13]$$

The appearance of the imaginary time direction also modifies the hyperscaling relations [7]: The spatial dimensionality $d$ has to be replaced by $d+z$. If space and time enter the theory symmetrically (as in the example [12]), the dynamical exponent is $z = 1$, but in general, it can take any positive value. Note that the quantum-to-classical mapping is valid for the thermodynamics only. Other properties like the real time dynamics at finite temperatures require more careful considerations. Moreover, the interpretation of the quantum partition function as a classical one in a higher dimension is only possible if the statistical weight is real and positive. If this is not the case (consider, e.g., Berry phase terms in the spin functional integral), the quantum-to-classical mapping cannot be applied directly.

The quantum-classical mapping can be exploited for computational studies of quantum phase transitions. If one is only interested in finding the universal critical behavior at the quantum critical point (i.e., in the critical exponents) and not in nonuniversal quantities, it is often easier to perform a simulation of the equivalent classical system instead of the original quantum system. We will come back to this point later in the chapter.

**Beyond the Landau-Ginzburg-Wilson paradigm**

In recent years, it has become clear that some quantum phase transitions cannot be satisfactorily described by the LGW approach, i.e., by considering long-wavelength fluctuations of a local order parameter only. In this section we briefly discuss mechanisms that can invalidate the LGW approach.

*Generic scale invariance.* The derivation of the LGW theory as a regular expansion of the free energy in terms of the order parameter fluctuations relies on these fluctuations being the only gapless (soft) modes in the system. If there are other soft modes, e.g., due to conservation laws or broken continuous symmetries, they lead to long-range power-law correlations of various quantities even away from the critical point. This phenomenon is called generic scale invariance.[32,33,34] If one insists on deriving a LGW theory in the presence of other gapless modes, the resulting functional has singular coefficients and is thus ill-defined. One should instead work with a coupled theory that keeps all soft modes at the same footing. This mechanism is discussed in detail in Ref. 9. It is important, e.g., for metallic quantum ferromagnets.

*Deconfined quantum criticality.* Certain two-dimensional $S=1/2$ quantum antiferromagnets can undergo a direct continuous quantum phase transition between two ordered phases, an antiferromagnetic Néel phase and the so-called valence-bond ordered phase (where translational invariance is broken). This is in contradiction to Landau theory which predicts phase coexistence, an intermediate phase, or a first-order transition, if any . The continuous transition is the result of topological defects that become spatially deconfined at the critical point and are not contained in a LGW description. Recently, there has been a great interest in the resulting deconfined quantum critical points. [35]

*Heavy-fermion quantum criticality.* Unconventional quantum critical point scenarios may be also important for understanding the magnetic transitions in heavy-fermion systems. In experiments,[36] many of these materials show pronounced deviations from the predictions of the standard LGW theory of metallic quantum phase transitions.[37,38] The breakdown of the conventional approach in these systems may have to do with the importance of Kondo fluctuations. The standard theory[37,38] assumes that the heavy quasiparticles (which are due to a Kondo hybridization between f and conduction electrons) remain intact across the transition. Other approaches start from the assumption that the Kondo effect breaks down right at the magnetic transition, a phenomenon which cannot be described in terms of the magnetic order parameter fluctuations. Several scenarios have been proposed, including the so-called local critical point,[39] and the fractionalized Fermi liquid leading to one of the above-mentioned deconfined quantum critical points.[40,41] At present, the correct theory for these transitions has not been worked out, yet.

**Impurity quantum phase transitions**

An interesting type of quantum phase transitions are boundary transitions where only the degrees of freedom of a subsystem become critical while the bulk remains uncritical. The simplest case is the so-called impurity quantum phase transitions where the free energy contribution of the impurity (or, in general, a zero-dimensional subsystem) becomes singular at the quantum critical point. Such transitions occur, e.g., in anisotropic Kondo systems, quantum dots, and in spin systems coupled to dissipative baths. A recent review can be found in Ref. 42. Impurity quantum phase transitions require the thermodynamic limit in the bulk (bath) system, but are completely independent from possible phase transitions of the bath.

# QUANTUM PHASE TRANSITIONS: COMPUTATIONAL CHALLENGES

Computational studies of quantum phase transitions generally require a very high numerical effort because they combine several formidable computational challenges. These include (i) the problem of many interacting degrees of freedom, (ii) the fact that phase transitions arise only in the thermodynamic limit of infinite system size, (iii) critical slowing down and supercritical slowing down at continuous and first-order transitions, respectively, and (iv) anisotropic space-time scaling at quantum critical points. In disordered systems, there is the additional complication (v) of having to simulate large numbers of disorder realizations to obtain averages and probability distributions of observables. In the following, we discuss these points in more detail.

(i) *The quantum many-particle problem.* At the core, computational studies of quantum phase transitions require simulating interacting quantum many-particle systems. The Hilbert space dimension of such systems increases exponentially with the number of degrees of freedom. Thus, "brute-force" methods such as exact diagonalization are limited to very small systems that are usually not sufficient to investigate the properties of phase transitions. In many areas of many-particle physics and chemistry, sophisticated approximation methods have been developed to overcome this problem. However, many of them are problematic in the context of quantum phase transitions. Selfconsistent-field (scf) or single-particle type approximations such as Hartree-Fock or density functional theory (see, e.g., Refs. 43-45) by construction neglect fluctuations because they express the many-particle interactions in terms of an effective field or potential. Since fluctuations have proven to be crucial for understanding continuous phase transitions (as discussed in the section on *Phase Transitions and Critical Behavior*), these methods must fail at least in describing the critical behavior close to the transition. They may be useful for approximately locating the transition in parameter space, though. Other approximation methods, such as the coupled cluster method,[46] go beyond the self-consistent field level by including one or several classes of fluctuations. However, since the set of fluctuations included is limited and has to be selected by hand, these methods are not bias-free. Quantum critical states are generally very far from any simple reference state; thus they are particularly challenging for these techniques. One important class of methods that are potentially numerically exact and bias-free are quantum Monte-Carlo methods.[47-49] They will be discussed in more detail later in this chapter. However, quantum Monte-Carlo methods for fermions suffer from the notorious sign-problem which originates in the antisymmetry of the many-fermion wave function and severely hampers the simulation. Techniques developed for dealing with the sign-problem often reintroduce biases into the method, for instance via forcing the nodes of the wave function to coincide with those of a trial wave function.

(ii) *Thermodynamic limit.* Sharp phase transitions only arise in the thermodynamic limit of infinite system size. Fortunately, this does not mean, one has to actually simulate infinitely large systems. The critical behavior of a continuous phase transition can be extracted from the simulation of finite systems by using finite-size scaling (see section on *Phase Transitions and Critical Behavior*). However, this still requires sufficiently large system sizes that are in the asymptotic finite-size scaling regime, where corrections to scaling forms such as eq. [8] are small. In general, it is not *a priori* clear how large the system sizes have to be to reach this asymptotic regime. Therefore, one must simulate a range of system sizes and test the validity of the scaling forms *a posteriori.*

(iii) *Critical and Supercritical Slowing Down.* As discussed in the section on *Phase Transitions and Critical Behavior,* critical points display the phenomenon of critical slowing down, i.e., the system dynamics becomes arbitrarily slow when one approaches the transition. First-order transitions can show an even more dramatic supercritical slowing down where the correlation time increases exponentially with the length scale. The same slowing down problem occurs in many Monte-Carlo methods, in particular if the updates (elementary moves) are local. This means that the necessary simulation times diverge when approaching the transition point. Critical and supercritical slowing down can be overcome by more sophisticated Monte-Carlo methods including cluster update techniques[50,51] for critical points and flat-histogram methods[52,53] for first-order transitions.

(iv) *Anisotropic space-time scaling at quantum critical points.* Many commonly used quantum Monte-Carlo algorithms work at finite temperatures and require an extrapolation to zero temperature for extracting information on quantum phase transitions. The data analysis in such simulations thus implies finite-size scaling not only for the spatial coordinates but also for the imaginary time direction. In general, this finite-size scaling will be anisotropic in space and time with an unknown dynamical exponent $z$. Therefore, system size and (inverse) temperature have to be varied independently, greatly increasing the numerical effort.

(v) *Disordered systems*. Computational studies of disordered systems in general require the simulation of a large number (from 100 to several 10000) of samples or disorder realizations to explore the averages or distribution functions of macroscopic observables. This is particularly important for finite-disorder and infinite-disorder critical points (which occur in many quantum systems) because at these critical points, the probability distributions of observables remain broad or even broaden without limit with increasing system size. Thus, the numerical effort for simulating a disordered quantum many-particle system can be several orders of magnitude larger than that for the corresponding clean system.

# CLASSICAL MONTE-CARLO APPROACHES

In this section, we describe computational approaches to quantum phase transitions that rely on the quantum-to-classical mapping. The number of transitions that can be studied by these approaches is huge; our discussion is therefore not meant to be comprehensive. After an introduction to the method we rather discuss a few characteristic examples, mostly from the area of magnetic quantum phase transitions.

**Method: quantum-to-classical mapping and classical Monte-Carlo methods**

As discussed in the section on *Quantum vs. Classical Phase Transitions,* the partition function of a $d$-dimensional quantum many-particle system can be written as a functional integral over space and (imaginary) time dependent fields. If the statistical weight in this representation is real and positive, it can be interpreted as the statistical weight of a classical system in $d+1$ dimensions with the extra dimension corresponding to the imaginary time direction. This classical system can now be simulated very efficiently using the well-developed machinery of classical Monte-Carlo methods.[54,55] Often, this quantum-to-classical mapping is exact only for the asymptotic low-energy degrees of freedom. Therefore, this approach works best if one is mostly interested in the universal critical behavior at the transition, i.e., in the overall scaling scenario and the values of the critical exponents, rather than nonuniversal quantities that depend on microscopic details such as the critical coupling constants or numerical values of observables..

In some particularly simple cases, the classical system arising from the quantum-to-classical mapping is in one of the well-known universality classes of classical phase transitions whose critical behavior has been studied in great detail in the literature. In these cases, the quantum problem can be solved by simply "translating" the known classical results and by calculating specific observables, if desired. The first two examples discussed below will be of this type. However, more often than not, the classical system arising from the quantum-to-classical mapping is unusual and anisotropic (space and imaginary time directions do not appear in a symmetric fashion). In these cases, the behavior of the classical system has likely not been studied before, but it can be simulated efficiently by classical Monte-Carlo methods.

**Transverse-field Ising model**

The first example is arguably the simplest model displaying a quantum phase transition, the quantum Ising model in a transverse field. It can be viewed as a toy model for the magnetic quantum phase transition of LiHoF$_4$ discussed in the introductory section. For this system, we now explain the quantum-to-classical mapping in detail, identify the equivalent classical model, and discuss the results for the quantum critical behavior.

The transverse field Ising model is defined on a *d*-dimensional hypercubic (i.e., square, cubic, etc.) lattice. Each site is occupied by a quantum spin-1/2. The spins interact via a ferromagnetic nearest-neighbor exchange interaction $J > 0$ between the *z*-components of the spins. The transverse magnetic field $h_x$ couples to the *x*-components of the spins. The Hamiltonian of the model is given by

$$\hat{H} = -J \sum_{\langle i,j \rangle} \hat{S}_i^z \hat{S}_j^z - h_x \sum_i \hat{S}_i^x \qquad [14]$$

For $J \gg h_x$, the system is in a ferromagnetic state, with a nonzero spontaneous magnetization in *z*-direction. In contrast, for $J \ll h_x$, the *z*-magnetization vanishes, and the system is a paramagnet. The two phases are separated by a quantum phase transition at $J \sim h_x$. The starting point for our investigation of the critical behavior of this transition is the partition function $Z = \text{Tr}\, e^{-\hat{H}/k_B T}$. We now show how to map this partition function onto that of a classical system. The procedure is analogous to Feynman's path integral for the quantum mechanical propagator.[56]

Because the $z$ and $x$ components of the spin operators do not commute, the partition function cannot be simply factorized into an interaction part and a transverse field part. However, we can use the Trotter product formula,[57] $e^{\hat{A}+\hat{B}} = \lim_{N\to\infty}\left(e^{\hat{A}/N}e^{\hat{B}/N}\right)^N$, for Hermitean operators $\hat{A}$ and $\hat{B}$ to slice the imaginary time direction and then factorize the exponential in each slice. The partition function now reads

$$Z = \text{Tr}\lim_{N\to\infty}\left(e^{\varepsilon J\sum_{\langle i,j\rangle}\hat{S}_i^z\hat{S}_j^z}\, e^{\varepsilon h_x\sum_i \hat{S}_i^x}\right)^N \qquad [15]$$

where $\varepsilon = 1/(k_B T N)$ is the step in imaginary time direction. We now insert resolutions of the unit operator in terms of $\hat{S}^z$ eigenstates between each pair of time slices as well as resolutions of the unit operator in terms of $\hat{S}^x$ eigenstates between the interaction and field terms within each slice. Applying all $\hat{S}^z$ operators onto $\hat{S}^z$ eigenstates and all $\hat{S}^x$ operators onto $\hat{S}^x$ eigenstates, we can express the partition function in terms of the eigenvalues (which are classical variables) only. The sums over the $\hat{S}^x$ eigenvalues can be easily carried out, and up to a constant prefactor, the partition function is given by

$$Z \propto \lim_{N\to\infty}\sum_{\{S_{i,n}\}} e^{\varepsilon J\sum_{\langle i,j\rangle,n} S_{i,n}S_{j,n} + K\sum_{i,n} S_{i,n}S_{i,n+1}} \qquad [16]$$

where $S_{i,n} = \pm 1$ is the $\hat{S}^z$ eigenvalue of the spin at site *i* and time slice *n*. The interaction K in imaginary time direction takes the form $K = \frac{1}{2}\ln\coth(\varepsilon h_x)$. This representation of the partition function of the transverse-field Ising model is identical to the partition function of an anisotropic classical Ising model in *d*+1 dimensions with coupling constants $\varepsilon J$ in the *d* space dimensions and *K* in the time-like direction. The classical Hamiltonian reads

$$H_{cl}/k_B T = -\varepsilon J \sum_{\langle i,j\rangle,n} S_{i,n} S_{j,n} - K \sum_{i,n} S_{i,n} S_{i,n+1} \qquad [17]$$

Since the interactions are short-ranged (nearest neighbor only) in both space and time-like directions, the anisotropy does not play a role for the critical behavior of this classical model. We thus conclude that the quantum phase transition of the *d*-dimensional quantum Ising model in a transverse field falls into the universality class of the (*d*+1)-dimensional classical Ising model. This establishes the quantum-to-classical mapping (for a slightly different derivation based on transfer matrices, see Ref. 10).

In this example, the classical model arising from the mapping is a well-studied model of classical statistical mechanics. We can thus simply translate the known results. Specifically, the one-dimensional transverse-field Ising model is equivalent to two-dimensional classical Ising model which was solved exactly in a seminal paper[58] by L. Onsager more than 60 years ago. The exponent values are $\alpha = 0, \beta = 1/8, \gamma = 7/4, \delta = 15, \nu = 1, \eta = 1/4$. Since space and time directions are equivalent, the dynamic exponent is $z = 1$. The critical behavior of various thermodynamic quantities can now be obtained from the homogeneity relation [13]. For instance, by differentiating [13] twice with respect to *h*, we obtain the homogeneity relation for the magnetic susceptibility

$$\chi(r,h,T) = b^{\gamma/\nu}\chi(rb^{1/\nu}, hb^{y_h}, Tb^z) \qquad [18]$$

Note that the field *h* appearing in [18] is a field conjugate to the order parameter, i.e., a magnetic field in *z* direction, not the transverse field $h_x$. By setting $r=0, h=0$ and $b=T^{-1/z}$, we find the temperature dependence of the zero-field susceptibility at criticality to be $\chi(T) \propto T^{-\gamma/\nu z} = T^{-7/4}$. Other thermodynamic observables can be determined analogously. The energy gap $\Delta$, an important property of the quantum system close to criticality, is related to the correlation length $\xi_t$ of the equivalent classical system in imaginary time direction via $\Delta^{-1} \propto \xi_t$.

The two-dimensional transverse-field Ising model maps onto the three-dimensional classical Ising model which is not exactly solvable. However, the critical behavior has been determined with high precision using Monte-Carlo and series expansion methods (see, e.g., Ref. 59). The exponent values are $\beta \approx 0.326, \gamma \approx 1.247, \nu \approx 0.629$. The other exponents can be found from the scaling and hyperscaling relations [6] and [7]. In dimensions three and higher, the transverse-field Ising model displays mean-field critical behavior because the equivalent classical model is at or above the upper critical dimension $d_c^+ = 4$. (Right at $d_c^+$, there will be the usual logarithmic corrections.[11])

**Bilayer Heisenberg quantum antiferromagnet**

A (single-layer) two-dimensional Heisenberg quantum antiferromagnet consists of quantum spins-1/2 on the sites of a square lattice. They interact via the Hamiltonian

$$\hat{H} = J_{\parallel} \sum_{\langle i,j \rangle} \hat{\mathbf{S}}_i \cdot \hat{\mathbf{S}}_j \qquad [19]$$

where $J_\parallel > 0$ is the nearest neighbor exchange interaction. In contrast to [14], the interaction is isotropic in spin space. This model describes, e.g., the magnetic properties of the CuO planes in undoped high-$T_c$ cuprate perovskites. Even though quantum fluctuations (caused by the non-commutativity of the spin components) reduce the staggered magnetization from its classical value 1/2 to about 0.3, the ground state displays long-range antiferromagnetic (Néel) order as will be discussed in the section on *Quantum Monte Carlo Methods*. In order to induce a quantum phase transition to a paramagnetic state, one has to increase the quantum fluctuations. This can be done, e.g., by considering two identical layers with the corresponding spins in the two layers coupled antiferromagnetically by an interaction $J_\perp > 0$ (see Fig. 4 for a sketch of the system).

The Hamiltonian of the resulting bilayer Heisenberg quantum antiferromagnet reads

$$\hat{H} = J_\parallel \sum_{\langle i,j \rangle} \left( \hat{\mathbf{S}}_{i,1} \cdot \hat{\mathbf{S}}_{j,1} + \hat{\mathbf{S}}_{i,2} \cdot \hat{\mathbf{S}}_{j,2} \right) + J_\perp \sum_i \hat{\mathbf{S}}_{i,1} \cdot \hat{\mathbf{S}}_{i,2} \qquad [20]$$

where the second index of the spin operator distinguishes the two layers. For $J_\perp \gg J_\parallel$, the corresponding spins in the two layers form singlets which are magnetically inert (i.e., $J_\perp$ increases the fluctuations away from the classical Néel state). Thus, the system is in the paramagnetic phase. In contrast, for $J_\perp \ll J_\parallel$, each layer orders antiferromagnetically, and the weak interlayer coupling establishes global antiferromagnetic long-range order. There is a quantum phase transition between the two phases at some $J_\perp \sim J_\parallel$.

We now map this quantum phase transition onto a classical one. Chakravarty, Halperin and Nelson[60] showed that the low-energy behavior of two-dimensional quantum Heisenberg antiferromagnets is generally described by a (2+1)-dimensional quantum rotor model with the Euclidean action

$$S = \frac{1}{2g} \int_0^{1/k_B T} d\tau \left[ \sum_i (\partial_\tau \mathbf{n}_i(\tau))^2 - \sum_{\langle i,j \rangle} \mathbf{n}_i(\tau) \cdot \mathbf{n}_j(\tau) \right] \quad [21]$$

or by the equivalent continuum nonlinear sigma model. Here $\mathbf{n}_i(\tau)$ is a three-dimensional unit vector representing the *staggered* magnetization. For the bilayer Hamiltonian [20], the rotor variable $\mathbf{n}_i(\tau)$ represents $\hat{\mathbf{S}}_{i,1} - \hat{\mathbf{S}}_{i,2}$ while the conjugate angular momentum represents $\hat{\mathbf{S}}_{i,1} + \hat{\mathbf{S}}_{i,2}$ (see chapter 5 of Ref. 10). The coupling constant $g$ is related to the ratio $J_\parallel / J_\perp$. By reinterpreting the imaginary time direction as additional space dimension we can now map the rotor model [21] onto a three-dimensional classical Heisenberg model with the Hamiltonian

$$H_{cl}/k_B T = -K \sum_{\langle i,j \rangle} \mathbf{n}_i \cdot \mathbf{n}_j \quad [22]$$

Here the value of $K$ is determined the ratio $J_\parallel / J_\perp$ and tunes the phase transition. (Since the interaction is short-ranged in space and time directions, the anisotropy of [21] does not play a role for the critical behavior.)

As in the first example, the classical system arising from the quantum-to-classical mapping is a well-known model of classical statistical mechanics. While it is not exactly solvable, its properties are known with high precision from classical Monte-Carlo simulations.[61,62] The critical exponents of the phase transition are $\alpha \approx -0.133, \beta \approx 0.369, \gamma \approx 1.396, \delta \approx 4.783, \nu \approx 0.711, \eta \approx 0.037$. Since space and time directions enter [22] symmetrically, the dynamical exponent is $z = 1$. The critical behavior of observables can be obtained from the homogeneity relation [13] as before. Note that the field $h$ appearing in the homogeneity relation is *not* a uniform magnetic field but rather the field conjugate to the antiferromagnetic order parameter, i.e., a staggered magnetic field. Including a uniform magnetic field in the quantum-to-classical mapping procedure leads to complex weights in the partition function. As a result, the uniform magnetic field has no analog in the classical problem. Investigating the effects of a uniform field beyond linear response theory is thus outside the quantum-to-classical mapping approach.

**Dissipative transverse-field Ising chain**

In the two examples above, the quantum-to-classical mapping resulted in systems where space and time directions appear symmetrically, implying a dynamical exponent $z=1$ from the outset. We now turn to an example where space and time directions scale differently; and the dynamical exponent has to be determined from the simulation data

The dissipative transverse-field Ising chain consists of a one-dimensional transverse-field Ising model as discussed in the first example with each spin coupled to a heat bath of harmonic oscillators. The Hamiltonian reads

$$\hat{H} = -J\sum_{\langle i,j \rangle}\hat{S}_i^z\hat{S}_j^z - h_x\sum_i \hat{S}_i^x + \sum_{i,k}\left[ c_k \hat{S}_i^z (a_{i,k}^\dagger + a_{i,k}) + \omega_{i,k} a_{i,k}^\dagger a_{i,k} \right] \quad [23]$$

Here $a_{i,k}^\dagger$ and $a_{i,k}$ are the creation and destruction operators of harmonic oscillator $k$ coupled to spin $i$. The oscillator frequencies $\omega_{i,k}$ and coupling constants $c_k$ are chosen such that the spectral function $J(\omega) = 4\pi \sum_k c_k^2 \delta(\omega - \omega_{i,k}) = 2\pi\alpha\omega$ for $\omega$ less than some cutoff $\omega_c$, but vanishes otherwise. This defines Ohmic (linear) dissipation with dimensionless dissipation strength $\alpha$.

The quantum-to-classical mapping for this system follows the same "Feynman path integral" procedure used in the first example. The harmonic oscillator degrees of freedom lead to Gaussian integrals and can thus be integrated out exactly. The resulting classical Hamiltonian reads

$$H_{cl}/k_B T = -\varepsilon J \sum_{\langle i,j \rangle,n} S_{i,n} S_{j,n} - K \sum_{i,n} S_{i,n} S_{i,n+1} - \frac{\alpha}{2} \sum_{i,n<m} \left(\frac{\pi}{N}\right)^2 \frac{S_{i,n} S_{i,m}}{\sin^2\left(\pi/N |n-m|\right)} \qquad [24]$$

Here $S_{i,n} = \pm 1$ are classical Ising variables at site $i$ and imaginary time step $n$. The time interval $\varepsilon$ is related to the inverse temperature via $\varepsilon = 1/(k_B T N)$, and the coupling constant $K$ is given by $K = \frac{1}{2} \ln \coth(\varepsilon h_x)$, as before. The coupling to the Ohmic baths has introduced a long-range interaction in time direction which behaves as $1/\tau^2$ in the Trotter limit $N \to \infty$. This long-range interaction breaks the symmetry between space and time directions.

The classical Hamiltonian [24] can now be studied using classical Monte Carlo algorithms. To reduce the effects of critical slowing down close to the transition, cluster algorithms are very desirable. However, the commonly used Swendsen-Wang[50] and Wolff[51] algorithms are not very efficient for long-range interactions because they have to go over all neighbors of each site when building a cluster. Luijten and Blöte[63] developed a version of the Swendsen-Wang algorithm that is suitable and efficient for long-range interactions. Werner et al.[64] used this algorithm to simulate the Hamiltonian [24]. Since space and time directions are not equivalent, the data analysis via finite-size scaling is not entirely trivial. An efficient way for determining the critical point and the dynamical exponent $z$ selfconsistently was suggested by Guo, Bhatt and Huse[65] as well as Rieger and Young.[66] It is based on analyzing dimensionless observables such as the Binder cumulant

$$B = 1 - \frac{\langle m^4 \rangle}{3\langle m^2 \rangle^2} \quad [25]$$

where *m* is the magnetization (i.e., the order parameter). This quantity approaches well-known limits in both bulk phases: In the ordered phase, all spins are correlated, and the magnetization has small fluctuations around a nonzero value. Therefore, $\langle m^4 \rangle \approx \langle m^2 \rangle^2$, and the Binder ratio approaches 2/3. In the disordered phase, the system consists of many independent fluctuators. Consequently, $\langle m^4 \rangle$ can be decomposed using Wick's theorem giving $\langle m^4 \rangle \approx 3\langle m^2 \rangle^2$, and the Binder ratio approaches zero. Since the Binder ratio is dimensionless, the finite-size scaling homogeneity relation for this quantity reads

$$B(r, L, L_t) = B(rb^{1/\nu}, Lb^{-1}, L_t b^{-z}) \quad [26]$$

where *L* and $L_t$ are the linear system sizes in space and time direction, respectively. Setting the arbitrary scale factor *b = L*, this leads to the scaling form

$$B(r, L, L_t) = \Phi_B(rL^{1/\nu}, L_t / L^z) \quad [27]$$

with $\Phi_B$ a dimensionless scaling function. An important characteristic follows: For fixed *L*, *B* has a peak as a function of $L_t$. The peak position $L_t^{max}$ marks the optimal sample shape, where the ratio $L_t / L$ roughly behaves like the corresponding ratio of the correlation lengths in time and space directions. (If the aspect ratio deviates from the optimal one, the system can be decomposed into independent units either in space or in time direction, and thus *B* decreases.) At the critical point, the peak value $B^{max}$ is independent of *L*. Thus, plotting *B* vs. $L_t / L_t^{max}$ at the critical point should collapse the data, without the need for a value of the dynamical exponent *z*. Instead, *z* can be extracted from the relation $L_t^{max} \propto L^z$. An example for such an analysis is shown in Fig 5. Once the dynamical exponent $z$ is found, the other exponents can be found from one-parameter finite-size scaling as in the classical case.[15,16,17]

Werner et al.[64] used these techniques to investigate the phase diagram of the Hamiltonian [24] and the quantum phase transition between the ferromagnetic and paramagnetic phases. They found the critical behavior to be universal (i.e., independent of the dissipation strength $\alpha$ for all $\alpha \ne 0$). The exponent values are $\nu \approx 0.638, \eta \approx 0.015, z \approx 1.985$. They agree well with the results of perturbative renormalization group calculations.[67,68] The other exponents can be found from the scaling and hyperscaling relations [6] and [7].

**Diluted bilayer quantum antiferromagnet**

In the last example, we have seen that dissipation can lead to an effective long-range interaction in time and thus break the symmetry between space and time directions. Another mechanism to break this symmetry is quenched disorder (i.e., impurities and defects), because it is random in space but perfectly correlated in time direction.

Consider for instance the bilayer Heisenberg quantum antiferromaget [20]. Random disorder can be introduced, e.g., by randomly removing spins from the system (in experiment, one would randomly replace magnetic atoms with nonmagnetic ones). If the substitutions in the two layers are made independently from each other, the resulting unpaired spins lead to complex weights in the partition function that cannot be expressed in terms of a classical Heisenberg model. Here, we therefore consider the case of dimmer-dilution, i.e., the corresponding spins in the two layers are removed together. The Hamiltonian of the dimer diluted bilayer Heisenberg quantum antiferromagnet is given by

$$\hat{H} = J_{\parallel} \sum_{\langle i,j \rangle} \mu_i \mu_j \left( \hat{\mathbf{S}}_{i,1} \cdot \hat{\mathbf{S}}_{j,1} + \hat{\mathbf{S}}_{i,2} \cdot \hat{\mathbf{S}}_{j,2} \right) + J_{\perp} \sum_i \mu_i \hat{\mathbf{S}}_{i,1} \cdot \hat{\mathbf{S}}_{i,2} \qquad [28]$$

where the $\mu_i$ are independent random variables that can take the values 0 and 1 with probability $p$ and $1-p$, respectively. The zero temperature phase diagram of this model has been worked out by Sandvik[69] and Vajk and Greven;[70] it is shown in Fig 6. For small $J_{\perp}$, magnetic long-range order survives up to the percolation threshold $p_p \approx 0.4072$ of the lattice, and a multicritical point exists at $J_{\perp}/J_{\parallel} \approx 0.16$, $p = p_p$. Thus, the dimer-diluted bilayer Heisenberg antiferromagnet has two quantum phase transitions, the generic transition for $p < p_p$ and a quantum percolation transition at $p = p_p$, $J_{\perp} < 0.16 J_{\parallel}$.

The quantum-to-classical mapping follows the same procedure as for the clean bilayer quantum Heisenberg model above. The result is an unusual diluted three-dimensional classical Heisenberg model. Because the impurities in the quantum system are quenched (time-independent), the equivalent classical Heisenberg model has line defects parallel to the imaginary time direction. The classical Hamiltonian is given by

$$H_{cl}/k_B T = -K \sum_{\langle i,j \rangle, n} \mu_i \mu_j \, \mathbf{n}_{i,n} \cdot \mathbf{n}_{j,n} - K \sum_{i,n} \mu_i \, \mathbf{n}_{i,n} \cdot \mathbf{n}_{i,n+1} \qquad [29]$$

where $i$ and $j$ are the spatial indices while $n$ is the index in the time-like direction. The line defects break the symmetry between space and time directions; we thus expect anisotropic scaling with a dynamical exponent $z \neq 1$.

Sknepnek, Vojta and Vojta[71] and Vojta and Sknepnek[72] have performed large scale Monte-Carlo simulations of the classical Hamiltonian [29] by means of the Wolff cluster algorithm.[51] Because of the disorder, all simulations involve averages over a large number (up to several 10,000) of disorder realizations. Let us first discuss the generic transition ($p < p_p$). As explained above, the scaling behavior of the Binder cumulant can be used to selfconsistently find the critical point and the dynamical exponent $z$. A typical result of these calculations is presented in Fig. 7. It shows the Binder cumulant at the critical point for a system with impurity concentration $p=1/5$. As can be seen in the main panel of this figure, the data scale very well when analyzed according to power-law scaling while the inset shows that they do not fulfill activated (exponential) scaling. Analogous data were obtained for impurity concentrations 1/8, 2/7 and 1/3. The dynamical exponent of the generic transition now follows from a power-law fit of the maximum position $L_t^{max}$ vs. $L$, as shown in Fig. 8. Taking the leading corrections to scaling into account, this gives a universal value $z \approx 1.31$. The correlation length exponent can be determined from the off-critical finite-size scaling of the binder cumulant, giving $\nu \approx 1.16$. Note that this value fulfills the inequality $d\nu > 2$ as required for a sharp transition in a disordered system (see discussion on quenched disorder in section *Phase Transitions and Critical Behavior*). Analyzing the magnetization and susceptibility data at criticality yields $\beta/\nu \approx 0.53, \gamma/\nu \approx 2.26$.

Vojta and Sknepnek[72] have also performed analogous calculations for the quantum percolation transition at $p = p_p, J_\perp < 0.16 J_\parallel$ and the multicritical point at $p = p_p, J_\perp = 0.16 J_\parallel$. A summary of the critical exponents for all three transitions is found in Table 3. The results for the percolation transition are in reasonable agreement with theoretical predictions of a recent general scaling theory[73] of percolation quantum phase transitions: $\beta/\nu = 5/48, \gamma/\nu = 59/16$ and a dynamical exponent of $z = D_f = 91/48$ (coinciding with the fractal dimension of the critical percolation cluster).

| Exponent | Generic Transition | Multicritical point | Percolation Transition |
|---|---|---|---|
| $z$ | 1.31 | 1.54 | 1.83 |
| $\beta/\nu$ | 0.53 | 0.40 | 0.15 |
| $\gamma/\nu$ | 2.26 | 2.71 | 3.51 |
| $\nu$ | 1.16 | | |

Table 3: Critical exponents of the generic transition, percolation transition and multicritical point of the dimer-diluted bilayer quantum Heisenberg antiferromagnet (from Ref. 72).

**Random transverse-field Ising model**

To illustrate the rich behavior of quantum phase transitions in disordered systems, we now consider the random transverse-field Ising model, a random version of our first example. It is given by the Hamiltonian

$$\hat{H} = -\sum_{\langle i,j \rangle} J_{ij}\, \hat{S}_i^z \hat{S}_j^z - \sum_i h_i^x\, \hat{S}_i^x \qquad [30]$$

where both $J_{ij} > 0$ and $h_i^x > 0$ are random functions of the lattice site. In one space dimension, the critical behavior of the quantum phase transition can be determined exactly[22,23] by means of the Ma-Dasgupta-Hu "strong-disorder" renormalization group.[24] This calculation predicts an exotic infinite-randomness critical point, characterized by the following unusual properties: (i) the effective disorder increases without limit under coarse graining (i.e. with increasing length scale), (ii) instead of the usual power law dynamical scaling one has activated scaling, $\ln \xi_t \propto \xi^\psi$, with $\psi = 1/2$, (iii) the probability distributions of observables become very broad, even on a logarithmic scale, with their widths diverging when approaching the critical point, (iv) average and typical correlations behave very differently: At criticality, the average correlations function $C_{av}(r)$ falls off with a power of the distance $r$, while the typical correlations decay much faster, as a stretched exponential $\ln C_{typ}(r) \propto r^{-\psi}$. These results have been confirmed by extensive efficient numerical simulations[74,75] based on mapping the spin systems onto free fermions.[76]

In dimensions $d > 1$, an exact solution is not available because the strong disorder renormalization group can be implemented only numerically.[19] Moreover, mapping the spin system onto free fermions is restricted to one dimension. Therefore, simulation studies have mostly used Monte-Carlo methods. The quantum-to-classical mapping for the Hamiltonian [30] can be performed analogously to the clean case. The result is a disordered classical Ising model in $d+1$ dimensions with the disorder perfectly correlated in one dimension (in 1+1 dimensions, this is the famous McCoy-Wu model[77,78]). The classical Hamiltonian reads

$$H_{cl}/k_B T = -\sum_{\langle i,j\rangle,n} (\varepsilon J_{ij}) S_{i,n} S_{j,n} - \sum_{i,n} K_i\, S_{i,n} S_{i,n+1} \qquad [31]$$

with $J_{ij} > 0$ and $K_i = \frac{1}{2} \ln \coth(\varepsilon h_i^x) > 0$ being independent random variables.

Pich et al.[79] performed Monte-Carlo simulations of this Hamiltonian in 2+1 dimensions using the Wolff cluster algorithm.[51] As in the two examples above, they used the scaling behavior of the Binder cumulant to find the critical point and to analyze the dynamical scaling. The resulting scaling plot is shown in Fig. 9. The figure shows that the curves do not scale when analyzed according to the usual power-law dynamical scaling, $\xi_t \propto \xi^z$, but rather get broader with increasing system size. In the inset, the data for $L \geq 12$ scale quite well according to activated scaling, $\ln \xi_t \propto \xi^\psi$, with $\psi \approx 0.42$. Pich et al.[79] also studied the behavior of the correlation function at criticality. They found a power-law decay of the average correlations and a stretched exponential decay of the typical correlations, as in one dimension. These results provide strong simulational evidence for the quantum critical point in the two-dimensional random transverse field Ising model being of exotic infinite randomness type. This agrees with the prediction of the numerically implemented strong-disorder renormalization group[19] and with a general classification of phase transitions in disordered systems according to the effective dimensionality of the defects.[31]

**Dirty bosons in two dimensions**

The examples discussed so far are all *magnetic* quantum phase transitions. Our last example in this section on quantum-to-classical mapping is a quite different transition, *viz.* the superconductor-insulator transition in two-dimensional dirty boson systems. Experimentally, this transition can be realized, e.g., in Helium absorbed in a porous medium or in granular superconducting films.

The minimal model for describing the superconductor-insulator transition in the general case of both charge and phase fluctuations being relevant is the boson Hubbard model with a random local chemical potential.[80,81] The Hamiltonian (defined on a square lattice) takes the form

$$\hat{H}_{BH} = \frac{U}{2} \sum_i \hat{N}_i^2 - \sum_i (\mu + v_i - zt) \hat{N}_i - t \sum_{\langle i,j \rangle} \left( \hat{\Phi}_i^\dagger \hat{\Phi}_j + \hat{\Phi}_j^\dagger \hat{\Phi}_i \right) \qquad [32]$$

Here, $U$ is the onsite repulsion, $\mu$ is the chemical potential, $z$ is the number of nearest neighbors and $v_i$ represents the random onsite potential. The hopping strength is given by $t$, and $\hat{\Phi}_i^\dagger$, $\hat{\Phi}_i$ are the boson creation and destruction operators at site $i$. The number operator is given by $\hat{N}_i = \hat{\Phi}_i^\dagger \hat{\Phi}_i$.

If the boson density is an integer (per site) and in the absence of disorder, charge (amplitude) fluctuations are small. If we set $\hat{\Phi}_i = |\hat{\Phi}_i| e^{i\hat{\theta}_i}$ and integrate out the amplitude fluctuations, we obtain a phase-only model that can be written as an O(2) quantum rotor model

$$\hat{H}_{QR} = -\frac{U}{2}\sum_i \frac{\partial^2}{\partial \theta_i^2} - t\sum_{\langle i,j\rangle} \cos(\theta_i - \theta_j) \quad [33]$$

This system describes, e.g., an array of coupled Josephson junctions.

In the spirit of this section, we now discuss the quantum-to-classical mapping for the dirty boson problem. We first consider the case of integer boson density and no disorder, i.e., the Hamiltonian [33]. In this case, the quantum-to-classical mapping can be performed analogously to the transverse-field Ising model: The partition function is factorized using the Trotter product formula leading to a path integral representation. By reinterpreting the imaginary time direction as an extra dimension and appropriately rescaling space and time (which does not change universal properties) we finally arrive at an isotropic three-dimensional classical XY model with the Hamiltonian

$$H_{cl}/k_B T = -K \sum_{\langle i,j\rangle} \cos(\theta_i - \theta_j) \quad [34]$$

where $\theta_i$ is a classical angle in the interval $[0, 2\pi]$. This is again a well-known model of classical statistical mechanics that can be simulated efficiently using Monte-Carlo cluster algorithms and series expansions (see, e.g., Ref. 82). The resulting critical exponents read $\alpha \approx -0.015$, $\beta \approx 0.348$, $\gamma \approx 1.318$, $\delta \approx 4.780$, $\nu \approx 0.672$, $\eta \approx 0.038$. Since space and time enter symmetrically, the dynamical exponent is $z = 1$.

The general case of noninteger boson density and/or the presence of the random potential is more realistic. However, it leads to broken time-reversal symmetry for the quantum rotors, because the particle number is represented by the quantity canonically conjugate to the phase variable, i.e., by angular momentum. The quantum-to-classical mapping procedure sketched above therefore leads to complex weights in the partition function, and the system cannot be interpreted in terms of a classical XY model. Wallin et al.[81] found an alternative quantum-to-classical mapping that avoids the complex weight problem. They expressed the partition function in terms of the integer-valued angular momentum variables of the rotors. The resulting link-current (Villain) representation is a classical (2+1)-dimensional Hamiltonian which reads

$$H_{cl}/k_B T = \frac{1}{K}\sum_{i,\tau}\left[\frac{1}{2}\mathbf{J}_{i,\tau}^2 - (\tilde{\mu}+\tilde{v}_i)J_{i,\tau}^\tau\right] \qquad [35]$$

Here, $i$ and $\tau$ are the site indices in space and the time-like direction, respectively. The dynamic variable $\mathbf{J} = (J^x, J^y, J^\tau)$ is a three-dimensional "current" with integer-valued components. It must be divergenceless, i.e., the sum over all currents entering a particular site must vanish. $\tilde{\mu}$ and $\tilde{v}_i$ represent the chemical and random potentials, renormalized by $U$.

To perform Monte-Carlo simulations of the classical Hamiltonian, one must construct updates that respect the zero divergence condition for the currents. This prevents the application of the usual type of cluster algorithms.[50,51] For this reason, early simulations[81] used algorithms with local updates which suffered from significant critical slowing down. Alet and Sorensen[83,84] developed a cluster algorithm in which the link currents are updated by moving a ''worm'' through the lattice. This algorithm is highly efficient and performs comparably to the Wolff algorithm[51] for classical spin systems. Using this algorithm, Alet and Sorensen first confirmed the three-dimensional XY universality class for the clean case at integer boson density. In the presence of the random potential, they found a different universality class with exponents $\nu \approx 1.15$ and $z \approx 2$.

# QUANTUM MONTE CARLO APPROACHES

If one is only interested in the universal critical behavior of a quantum phase transition, then the quantum-to-classical mapping method discussed in the last section (if available) is usually the most efficient approach. However, if one is also interested in nonuniversal quantities such as critical coupling constants or numerical values of observables, the quantum system has to be simulated directly. This can be done, e.g., by quantum Monte Carlo methods which are the topic of this section.

The name *quantum Monte Carlo* refers to a diverse class of algorithms used for simulating quantum many-particle systems by stochastic means (for an overview see, e.g., Ref. 47). Some of these algorithms such as variational Monte Carlo[85,86] and diffusion Monte Carlo[87,88] aim at computing the ground state wave function (and are thus zero-temperature methods). Other algorithms including path-integral (world-line) Monte Carlo[89,90] sample the density matrix at finite temperatures. Before we dive into discussing quantum phase transitions, it is useful to illustrate the wide spectrum of problems that can be attacked by quantum Monte Carlo methods today and the different challenges involved.

One branch of quantum Monte Carlo research aims at a quantitative first-principle description of atoms, molecules and solids beyond the accuracy provided by density functional theory.[48,49] If the basic physics and chemistry of the material in question is well understood at least *qualitatively* (as is the case, e.g., for many bulk semiconductors), good trial wave functions, e.g., of Jastrow-Slater type can be constructed. They can then be used in variational or diffusion

Monte Carlo simulations to provide high accuracy results for the correlation energy and other quantities. In contrast, materials whose behavior is not even qualitatively understood, such as many strongly correlated electron systems, pose different problems. They are often studied via simple models that capture the new properties of a whole class of materials without adding too many (realistic) details. However, the absence of even a qualitative understanding severely hampers the construction of trial wave functions with the right properties (symmetries etc.). Ideally, this class of problems is therefore studied by (bias-free) methods that do not rely on trial wave functions at all.

Simulating quantum phase transitions belongs squarely to the second class of problems. While variational or diffusion Monte Carlo calculations can be very useful in approximately locating the quantum phase transition of a particular system in parameter space, they are much less suitable for studying the quantum critical state itself because it is generally far away from any simple reference state. In recent years, significant progress in simulating quantum phase transitions of boson and spin systems has been achieved by path-integral (world-line) Monte Carlo[89,90] and the related stochastic series expansion (SSE) method.[91,92] Fermion systems pose a much harder problem because the antisymmetry of the many-fermion wave function generically leads to the notorious sign-problem. We will come back to this case at the end of the section. In the following we briefly introduce the world-line and SSE methods and then discuss a few characteristic examples of quantum phase transitions in boson and spin systems.

**World-line Monte Carlo**

The world-line Monte Carlo algorithm is a finite-temperature method that samples the canonical density matrix of a quantum many-particle system. At first glance, it may appear counterintuitive to use a finite-temperature method to study quantum phase transitions which occur at zero temperature. However, this is incorrect for at least the following two reasons: (i) One of the (experimentally) most interesting regions of the phase diagram close to a quantum critical point is the quantum critical region located at the critical coupling strength but comparatively high temperatures (see section on *Quantum vs. Classical Phase Transitions*). Finite-temperature methods are thus required to explore it. (ii) The dependence of observables on temperature is a very efficient tool for determining the dynamical scaling behavior of the quantum critical point (analogously to finite-size scaling, but in imaginary time direction).

The general idea[89,90] of the world-line Monte Carlo algorithm is very similar to that of the quantum-to-classical mapping discussed in the last section. The Hamiltonian is split into two or more terms $\hat{H} = \sum_i \hat{H}_i$ such that the matrix elements of each exponential term $e^{-\varepsilon \hat{H}_i}$ can be easily calculated. Even if the $\hat{H}_i$ do not commute, we can use the Trotter product formula to decompose the canonical density operator

$$e^{-\hat{H}/k_B T} = \lim_{N \to \infty} \left( \prod_i e^{-\varepsilon \hat{H}_i} \right)^N \qquad [36]$$

with $\varepsilon = 1/k_B T N$. Inserting complete sets of states between the different factors leads to a representation of the Boltzmann factor in terms of matrix elements of the $e^{-\varepsilon \hat{H}_i}$. If all these matrix elements are positive, their product can be interpreted as statistical weight, and Monte Carlo algorithms to sample this weight can be constructed. (If some of the matrix elements are negative, we have an instance of the notorious sign problem in quantum Monte Carlo.) The *N* factors of the Trotter decomposition can be interpreted as *N* time slices in imaginary time direction with particles or spins moving on "world lines" in the ($d$+1)-dimensional space-time. This gives the method its name. A specific implementation of the world-line Monte Carlo method will be discussed in our first example further down. More details can also be found, e.g., in chapter 3 of Ref. 47.

Applications of the world-line algorithm to quantum phase transitions require three extrapolations: (i) infinite system size, (ii) temperature $T \to 0$, and (iii) imaginary time step $\varepsilon \to 0$. The first two extrapolations can be handled conveniently by finite-size scaling in the space and time directions, respectively. The systematic error of the Trotter decomposition arising from a finite $\varepsilon$ was originally controlled by an explicit extrapolation from simulations with different values of $\varepsilon$. In 1996, Prokofev et al. showed that (at least for quantum lattice models) the algorithm can be formulated in continuous time, taking the limit $\varepsilon \to 0$ from the outset.[93] World-line Monte Carlo algorithm with local updates of the spin or particle configurations suffer from critical slowing down close to quantum critical points. This problem is overcome by the loop algorithm[94] and its continuous time version.[95] These algorithms which are generalizations of the classical cluster algorithms[50,51] to the quantum case, have been reviewed in detail in Ref. 96. Further improvements for systems without spin-inversion or particle-hole symmetry include the worm algorithm[97] and the directed loop method.[98]

**Stochastic series expansion**

The stochastic series expansion (SSE) algorithm[91,92] is a generalization of Handscomb's power-series method[99] for the Heisenberg model. To derive a SSE representation of the partition function, we start from a Taylor expansion in powers of the inverse temperature. We then decompose the Hamiltonian into two or more terms $\hat{H} = \sum_i \hat{H}_i$ such that the matrix elements with respect to some basis can be easily calculated, giving

$$Z = \mathrm{Tr}\, e^{-\hat{H}/k_B T} = \sum_{n=0}^{\infty} \frac{1}{n!}\left(\frac{1}{k_B T}\right)^n \mathrm{Tr}\left(-\hat{H}\right)^n = \sum_{n=0}^{\infty} \frac{1}{n!}\left(\frac{1}{k_B T}\right)^n \mathrm{Tr}\left(-\sum_i \hat{H}_i\right)^n \qquad [37]$$

Inserting complete sets of basis states between the different $\hat{H}_i$ factors then leads to a similar representation of the partition function and a similar world-line picture as in the world-line Monte Carlo method. Since there is no Trotter decomposition involved, the method is free of time discretization errors from the outset. Early applications of the SSE method employed local updates, but more recently, much more efficient cluster-type updates have been developed to overcome the critical slowing down. They include the operator-loop update[100] and the already mentioned directed loop algorithm.[98]

The source code for some of the algorithms discussed above is available on the WWW as part of the ALPS (Algorithms and Libraries for Physics Simulations) project.[101] SSE programs for the Heisenberg model can also be found on the homepage of A. Sandvik.[102]

**Spin-1/2 quantum Heisenberg magnet**

We will use our first example, the spin-1/2 quantum Heisenberg magnet, to further illustrate the world-line quantum Monte Carlo method. The model we study is the quantum XXZ model

$$\hat{H} = \sum_{\langle i,j \rangle} \left[ J_x \left( \hat{S}_i^x \hat{S}_j^x + \hat{S}_i^y \hat{S}_j^y \right) + J_z \hat{S}_i^z \hat{S}_j^z \right] = \sum_{\langle i,j \rangle} \left[ \frac{J_x}{2} \left( \hat{S}_i^+ \hat{S}_j^- + \hat{S}_i^- \hat{S}_j^+ \right) + J_z \hat{S}_i^z \hat{S}_j^z \right] \qquad [38]$$

where $\hat{S}_i^x, \hat{S}_i^y, \hat{S}_i^z$ are the components of the quantum spin-1/2 operator at site $i$, $\hat{S}_i^+, \hat{S}_i^-$ are the associate raising and lowering operators, and the sum is over all pairs of nearest neighbors. We now divide the Hamiltonian into pieces such that the matrix elements of each piece can be easily evaluated. For the XXZ Hamiltonian, a convenient choice is the so-called checkerboard decomposition.[103] Let us illustrate it by considering one space dimension (see also Fig. 10). We write $\hat{H} = \hat{H}_1 + \hat{H}_2$ where $\hat{H}_1$ contains the bonds between sites $i$ and $i+1$ for all even $i$ while $\hat{H}_2$ contains those for all odd $i$. To find a world-line representation we now insert this decomposition into the Trotter formula [36]. Since $\hat{H}_1$ and $\hat{H}_2$ each consist of independent two-site terms, the matrix elements in a $\hat{S}^z$ basis of $e^{-\varepsilon \hat{H}_i}$ completely factorize into terms of the type

$$e^{-\varepsilon J_z S^z_{i,n} S^z_{i+1,n}} \left\langle S^z_{i,n} S^z_{i+,n} \left| e^{-\varepsilon \frac{J_x}{2}\left(\hat{S}^+_i \hat{S}^-_{i+1} + \hat{S}^-_i \hat{S}^+_{i+1}\right)} \right| S^z_{i,n+1} S^z_{i+,n+1} \right\rangle \quad [39]$$

where $n$ is the Trotter index. The remaining matrix elements are easily calculated. They read (with $\hat{h} = J_x \left(\hat{S}^+_i \hat{S}^-_{i+1} + \hat{S}^-_i \hat{S}^+_{i+1}\right)/2$)

$$\begin{aligned}
\langle ++|e^{-\varepsilon \hat{h}}|++\rangle &= \langle --|e^{-\varepsilon \hat{h}}|--\rangle = 1 \\
\langle +-|e^{-\varepsilon \hat{h}}|+-\rangle &= \langle -+|e^{-\varepsilon \hat{h}}|-+\rangle = \cosh(\varepsilon J_x/2) \\
\langle +-|e^{-\varepsilon \hat{h}}|-+\rangle &= \langle -+|e^{-\varepsilon \hat{h}}|+-\rangle = -\sinh(\varepsilon J_x/2)
\end{aligned} \quad [40]$$

All other matrix elements are zero. The only non-vanishing matrix elements are those between states with the same total spin in the two Trotter slices, reflecting the spin conservation of the Hamiltonian. Note that the off-diagonal matrix elements are negative if $J_x$ is antiferromagnetic ($J_x > 0$). This prevents interpreting the matrix elements as statistical weight and indicates an instance of the sign problem. However, for our one-dimensional chain, or more general, on any bipartite lattice, we can eliminate the sign problem by rotating every other spin by 180 degrees which changes the sign of $J_x$.

The allowed spin configurations can be easily visualized in a (1+1)-dimensional space-time picture by drawing lines connecting space-time points where the z-component of the spin points up (see Fig. 11). Since the number of such sites is conserved, the resulting "world lines" are continuous. Moreover, the periodic boundary conditions implied by the trace require that the world lines also connect continuously from the last imaginary time slice to the first.

As the last ingredient for the Monte Carlo algorithm, we have to specify the Monte Carlo moves within the restricted class of allowed spin configurations. Single spin flips are not allowed, as they break the continuous world lines. Instead, the simplest Monte Carlo moves consist in proposing a local deformation of the world line (an example is shown in Fig. 11) and accepting or rejecting it with a suitable (Metropolis) probability determined by the changes in the matrix elements involved. As discussed above, algorithms based on such local moves suffer from critical slowing down near a quantum critical point. In the more efficient loop algorithm,[94,95,96] one builds large world line loops and then changes the spin direction along the entire loop.

Let us now focus on the special case of the isotropic ($J_x = J_z > 0$) Heisenberg quantum antiferromagnet on the square lattice (see also Hamiltonian [19]). This model has played an important role in the history of high-temperature superconductivity because it describes the magnetic properties of the copper oxide planes in the undoped parent cuprate perovskites. An important early problem was establishing beyond doubt that the ground state is antiferromagnetically (Néel) ordered and finding the value of the staggered magnetization. Reger and Young[104] performed world-line Monte Carlo simulations of the square lattice Heisenberg antiferromagnet using a two-dimensional version of the algorithm described above. Since the ground state of the Heisenberg model for any finite system size is rotationally invariant, the expectation value of the staggered magnetization vanishes. To determine the macroscopic value which assumes that the symmetry has been spontaneously broken, Reger and Young computed both the (staggered) structure factor $S(\mathbf{Q})$ at the ordering wave vector $\mathbf{Q} = (\pi, \pi)$ and the correlation function $C_{L/2}$ between spins as far apart as possible on the lattice. In the thermodynamic limit, both quantities reduce to $m_s^2/3$ where $m_s$ is the staggered magnetization. Fig. 12 shows the extrapolation of $S(\mathbf{Q})$ and $C_{L/2}$ to infinite system size (the extrapolations $T \to 0$ and $\varepsilon \to 0$ have already been carried out). From the intercept with the vertical axis, Reger and Young found $m_s = 0.30 \pm 0.02$ clearly establishing that the ground state is antiferromagnetically ordered. In later work, the staggered magnetization value was further refined by simulations using a continuous time loop algorithm,[95] giving the value $m_s = 0.3083 \pm 0.0002$.

**Bilayer Heisenberg quantum antiferromagnet**

While quantum fluctuations reduce the staggered magnetization of a single-layer Heisenberg quantum antiferromagnet from its classical value of ½, they are not strong enough to induce a quantum phase transition. As discussed in the section on *Classical Monte Carlo Approaches*, the strength of the quantum fluctuations can be tuned if one considers a system of two identical, antiferromagnetically coupled layers defined by the bilayer Hamiltonian [20]. If the interlayer coupling $J_\perp$ is large compared to the in-plane coupling $J_\parallel$, the corresponding spins in the two layers form magnetically inert singlets. In contrast, for $J_\perp \ll J_\parallel$, the system orders antiferromagnetically. There is a quantum phase transition between these two phases at some critical value of the ratio $J_\perp / J_\parallel$.

In the section on *Classical Monte Carlo Approaches* we have used the quantum-to-classical mapping to discuss the universal critical behavior of this quantum phase transition and found it to be in the three-dimensional classical Heisenberg universality class. However, this approach does not give quantitative answers for non-universal observables such as the critical value of the ratio $J_\perp / J_\parallel$ which can only be obtained by a true quantum algorithm. Sandvik and Scalapino[105] have performed quantum Monte-Carlo simulations of the bilayer Heisenberg quantum antiferromagnet employing the stochastic series expansion method. By analyzing the staggered structure factor and the staggered susceptibility they found a critical ratio of $(J_\perp / J_\parallel)_c = 2.51 \pm 0.02$ (see the vertical axis in Fig. 6). Very recently, Wang et al.[106] performed a high-precision study of the same model using the stochastic series expansion algorithm with operator loop update.[100] Using the Binder cumulant, the spin stiffness and the uniform susceptibility, they obtained a value of $(J_\perp / J_\parallel)_c = 2.5220 \pm 0.0001$ for the critical coupling. In addition, they also computed the correlation length exponent and found $\nu = 0.7106 \pm 0.0009$ which agrees within error bars with the best value of the three-dimensional classical Heisenberg exponent[62] (as expected from the quantum-to-classical mapping).

**Diluted Heisenberg magnets**

In the example above, we have seen that increased quantum fluctuations (as induced by the inter-layer coupling $J_\perp$ in the bilayer system) can cause a quantum phase transition in the two-dimensional Heisenberg quantum antiferromagnet. Another way to increase the fluctuations is by dilution, i.e., by randomly removing spins from the lattice. The phases and phase transitions of diluted Heisenberg quantum antiferromagnets have been studied extensively during the last few years and many interesting features have emerged.

Consider, e.g., the site diluted square lattice Heisenberg model given by the Hamiltonian

$$\hat{H} = J \sum_{\langle i,j \rangle} \mu_i \mu_j \, \hat{\mathbf{S}}_i \cdot \hat{\mathbf{S}}_j \qquad [41]$$

where the $\mu_i$ are independent random variables that can take the values 0 and 1 with probability $p$ and 1-$p$, respectively. As discussed above, the ground state of the clean system ($p = 0$) is aniferromagnetically ordered. It is clear that the tendency towards magnetism decreases with increasing impurity concentration $p$, but the location and nature of the phase transition towards a nonmagnetic ground state was controversial for a long time. The most basic question is whether the magnetic order vanishes before the impurity concentration reaches the percolation threshold $p_p \approx 0.4072$ of the lattice (the transition would then be caused by quantum fluctuations) or whether it survives up to $p_p$ (in which case the transition would be of percolation type). Magnetic long-range order is impossible above $p_p$, because the lattice is decomposed into disjoined finite-size clusters. Various early studies, both analytical and numerical, gave values between 0.07 and 0.35 for the critical impurity concentration, suggesting a transition driven by quantum fluctuations.

Sandvik[107] performed quantum Monte Carlo simulations of the Heisenberg Hamiltonian on the critical infinite percolation cluster ($p = p_p$) using the stochastic series expansion method with operator loop update.[100] He computed the staggered structure factor and from it the staggered ground state magnetization of the cluster. Figure 13 shows the extrapolation of this quantity to infinite system size. The data clearly demonstrate that the ground state is magnetically ordered, with a sizable staggered magnetization of about $m_s = 0.150$ (roughly half the value of the undiluted system). This means, even right at the percolation threshold $p_p$, the quantum fluctuations are not strong enough to destroy the magnetic long-range order. The phase transition to a paramagnetic ground state occurs right at $p_p$. It is driven by the geometry of the underlying lattice and thus of percolation type. More recently, Wang and Sandvik[108] studied the dynamical quantum critical behavior of this transition (the static one is given by classical percolation). They found a dynamical critical exponent of $z \approx 3.7$, much larger than the value $z = D_f = 91/48$ found for the dimer diluted bilayer[72,73] (see discussion in the section on *Classical Monte Carlo Approaches*). This difference is most likely caused by unpaired spins (uncompensated Berry phases) that exist in the site-diluted single layer (but not in the dimer-diluted bilayer) and prevent the quantum-to-classical mapping onto a classical Heisenberg model.

Since the ground state of the diluted Heisenberg model remains long-range ordered up to the percolation threshold, one has to increase the quantum fluctuations to induce a quantum phase transition for $p < p_p$. One way to achieve this is by going to the (dimer-diluted) bilayer (as in the clean system) and tuning the fluctuations with the interlayer coupling $J_\perp$. The quantum phase transitions in this system have been discussed in the section on *Classical Monte Carlo Approaches* above. Yu et al.[109] found a different way of increasing the quantum fluctuations. They suggested introducing *inhomogeneous* bond dilution, i.e., not all bonds (interactions) are removed with the same probability. If the occupation probabilities for different types of bonds are chosen in such a way that the system preferably forms dimers and ladders, a nontrivial quantum phase transition to a paramagnetic ground state can be achieved while the underlying lattice is still in the percolating phase.

**Superfluid-insulator transition in an optical lattice**

After having considered several examples of magnetic quantum phase transitions, we now turn to the superfluid-insulator transition in many-boson systems. In the section on *Classical Monte Carlo Approaches* we have discussed how the universal critical behavior of this transition can be determined by mapping the Bose-Hubbard model [32] onto the classical ($d$+1) dimensional link-current Hamiltonian [35] which can then be simulated using classical Monte Carlo methods.

In recent years, it has become possible to experimentally observe this transition in ultracold atomic gases. For instance, in the experiments by Greiner et al.,[110] a gas of $^{87}$Rb atoms is trapped in a simple cubic optical lattice potential. This system is well described by the Bose-Hubbard Hamiltonian [32] with an additional overall harmonic confining potential; and the particles density as well as the interparticle interactions can be easily controlled. To study the properties of the gas in the experiment, the trapping and lattice potential are switched off, and absorption images of the freely evolving atomic cloud are taken. This gives direct information about the single-particle momentum distribution of the gas.

To provide quantitative predictions as to how to detect the superfluid-insulator transition in these experiments, Kashurnikov et al.[111] performed quantum Monte Carlo simulations of the single particle density matrix $\rho_{ij} = \langle \hat{\Phi}_i^\dagger \hat{\Phi}_j \rangle$ of the Bose Hubbard model with harmonic confining potential using world-line Monte Carlo simulations with the continuous-time Worm algorithm.[97] The diagonal elements of the density matrix give the real-space particle density, and the momentum distribution can be obtained by Fourier transforming $\rho_{ij}$. The real-space density of several example systems is shown in Fig. 14. It features a shell-type structure with insulator phases visible as plateaus at integer local density. Specifically, system (a) is in the superfluid phase. If the Hubbard interaction $U$ is raised slightly above the critical value of the superfluid-insulator transition (system (b)), an insulating domain appears at the center of the trap (if the density there is close to commensurate). Increasing $U$ further reduces the correlation length (system (c)) because the system moves away from the transition. Systems (d) to (f) show how the shell structure develops when the density is increased.

The corresponding momentum distributions are shown in Fig. 15. The superfluid sample (a) shows a single narrow peak at zero momentum. (The broadening of the $\delta$-function contribution of the condensate expected in a superfluid is due to the harmonic confining potential.) When a domain of the insulating phase appears, the momentum distribution develops a pronounced fine structure (systems (b), (c), (d)). System (e) is similar to (a) except for the large momentum tail due to the insulating shell. System (f) again displays the fine structure associate with the appearance of an insulating domain in the second shell. These quantitative results can be used to identify the superfluid-insulator transition in experiments.

**Fermions**

So far, we have discussed quantum Monte Carlo approaches to quantum phase transitions in boson and spin systems. In these systems, the quantum Monte Carlo methods generically do not have a sign problem, i.e., the statistical weight in the Monte Carlo procedure this positive definite. Note that for spin systems, this is only true if there is no frustration. Frustrated spin systems in general do have a sign problem.

Unfortunately, the sign problem is generic for fermions because it is rooted in the antisymmetry of the many-fermion wave function. This can be understood as follows: boson and spin operators on different lattice sites commute. The signs of the matrix elements appearing in a quantum Monte Carlo scheme are thus determined locally. In contrast, fermion operators on different lattice sites anticommute leading to extra nonlocal minus signs. In fact, it was recently shown that a generic solution to the sign problem is almost certainly impossible by proving that the sign problem belongs to the NP (nondeterministic polynomial) hard computational complexity class.[112] This means that a generic solution of the sign problem would also solve all other NP hard problems in polynomial time.

One way of circumventing (if not solving) the sign problem consists in forcing the nodes of the many-fermion wave function to coincide with that of a trial wave function. The resulting fixed-node quantum Monte Carlo method[88,113] has been very successful in determining the ground state properties of real materials with high precision. It is clear that the accuracy of the method crucially depends on the quality of the trial wave function. This implies that it will work rather well if the ground state properties are at least qualitatively well understood. However, quantum critical states are in general very far from any simple reference state; and simple trial wave functions cannot easily be constructed. This makes fixed-node methods not very suitable for studying the properties of fermionic systems close to quantum phase transitions (although they may be useful for locating the transition in parameter space).

While a general solution of the fermionic sign problem is likely impossible, there are several nontrivial fermionic systems for which the sign problem can be avoided. Hirsch et al.[90] developed a world-line Monte Carlo simulation scheme for fermions. In strictly one dimension, this method avoids the sign problem, but generalizations to higher dimensions generically suffer from it. A more general approach is the so-called determinantal Monte Carlo method.[114] Its basic idea is to decouple the fermion-fermion interactions by means of a Hubbard-Stratonovich transformation,[115] leading to a system of noninteracting fermions coupled to a bosonic field. The fermions can now be integrated out in closed form, and the partition function is given as the sum over configurations of the bosonic field with the weight being a fermionic determinant. This sum can be performed by Monte Carlo sampling. In general, the fermionic determinant will have a fluctuating sign, again reflecting the fermionic sign problem. However, in some special cases the determinant can be shown to be positive definite. For instance, for the two-dimensional repulsive Hubbard model on bipartite lattices, the determinant is positive definite at exactly half filling (because of particle-hole symmetry).[116] For the attractive Hubbard model, sign-problem free algorithms can even be constructed for all filling factors. These algorithms have been used to study the superconducting transition in two and three spatial dimensions. In two dimensions, the transition is of Kosterlitz-Thouless type.[117,118,119] In three dimensions, the model displays a conventional second-order transition and an interesting crossover between the Bardeen-Cooper-Schrieffer (BCS) and Bose-Einstein condensation (BEC) scenarios.[120] Another attack on the sign problem is by the so-called meron-cluster algorithm that can be applied to certain fermionic

Hamiltonians.[121] It has been used, e.g., to study the effects of disorder superconductivity in fermion models with attractive interactions.[122]

Despite this progress, the utility of quantum Monte Carlo simulations in studying quantum phase transitions in fermionic systems is still rather limited. Many of the most interesting problems, such as the ferromagnetic and antiferromagnetic quantum phase transitions[9,36,123] in transition metal compounds and heavy-fermion materials are still far too complex to be directly attacked by microscopic quantum Monte Carlo methods.

# OTHER METHODS AND TECHNIQUES

In this section we briefly discuss – without any pretense of completeness - further computational approaches to quantum phase transitions. The conceptually simplest method for solving a quantum many-particle problem is arguably (numerically) exact diagonalization. However, as already discussed in the section on *Computational Challenges,* the exponential increase of the Hilbert space dimension with the number of degrees of freedom severely limits the possible system sizes. Even for simple lattice systems, one can rarely simulate more than a few dozen particles. Generally, these sizes are too small to study quantum phase transitions (which are a property of the thermodynamic limit of infinite system size) maybe with the exception of certain simple one-dimensional systems. However, in one dimension, more powerful methods have largely superceded exact diagonalization.

One of these techniques is the density matrix renormalization group (DMRG) proposed by White in 1992.[124] In this method, one builds the eigenstates of a large many-particle system iteratively from the low-energy states of smaller blocks, using the density matrix to decide which states to keep and which to discard. In one space dimension, this procedure works very well and gives accuracies comparable to exact diagonalization, but for much larger system sizes. Since its introduction, the DMRG has quickly become a method of choice for many one-dimensional quantum many-particle problems including various spin chains and spin ladders with and without frustration. Electronic systems such as Hubbard chains and Hubbard ladders can be studied efficiently as well because the DMRG is free of the fermionic sign problem. An extensive review of the DMRG method and its applications can be found in Ref. 125.

However, in the context of our interest in quantum phase transitions, it must be pointed out that the accuracy of the DMRG method greatly suffers in the vicinity of quantum critical points. This was shown explicitly in two studies of the one-dimensional Ising model in a transverse field, as given by the Hamiltonian [14].[126,127] Legaza and Fath[127] studied chains of up to 300 sites and found that the relative error of the ground state energy at the quantum critical point is several orders of magnitude larger than off criticality. (This is caused by the fact that the quantum critical system is gapless; it thus has many low-energy excitations that have to be kept in the procedure.) Andersson, Boman and Östlund[128] studied the behavior of the correlation function in a DMRG study of gapless free fermions (or equivalently, a spin-1/2 XX model). They found that the DMRG result reproduces the correct power law at small distances but always drops exponentially at large distances. The fake correlation length grows as $M^{1.3}$ with the number of states $M$ retained in each DMRG step. When studying a critical point, this fake correlation length should be larger than the physical correlation length which greatly increases the numerical effort. While the standard DMRG method does not work very well in dimensions larger than one, recently, an interesting generalization[129,130] has arisen in the quantum information community. It is based on so-called projected entangled-pair states (PEPS). First applications to quantum many-particle systems look promising (e.g., Ref 131 for a study of bosons in a two-dimensional optical lattice), but the true power of the method has not been fully explored.

Another very useful technique for studying one-dimensional spin systems is mapping the system onto *noninteracting* fermions. This method was developed by Lieb, Schultz and Mattis[76] in the 1960's and applied to the nonrandom transverse-field Ising model [14] by Katsura[132] and Pfeuty.[133] In the nonrandom case, the resulting fermionic Hamiltonian can be solved analytically by Fourier transformation. Young and Rieger[74,75] applied the same method to the random transverse-field Ising chain [30]. The mapping onto fermions now results in a disordered system; the fermionic Hamiltonian must therefore be diagonalized numerically. However, since one is simulating a *noninteracting* system, the numerical effort is still much smaller than with other methods. Using this approach, Young and Rieger numerically confirmed the analytical result[22,23] that the quantum critical point in the random transverse-field Ising chain is of exotic infinite-randomness type.

In recent years, the investigation of quantum phase transitions in disordered systems has strongly benefited from the strong-disorder renormalization group which was originally introduced by Ma, Dasgupta, and Hu.[24] The basic idea of this method is to successively integrate out local high-energy degrees of freedom in perturbation theory. In contrast to many other techniques, the quality of this method improves with increasing disorder strength; and it becomes asymptotically exact at infinite-randomness critical points (where the effective disorder strength diverges). By now, this approach has been applied to a variety of classical and quantum disordered systems, ranging from quantum spin chains to chemical reaction-diffusion models with disorder. A recent review can be found in Ref. 134. In one space dimension, the strong-disorder renormalization group can often be solved analytically in closed form, as is the case, e.g., for the random transverse-field Ising chain[22,23] or the random S=1/2 antiferromagnetic Heisenberg chain.[24,135] In higher dimensions, or for more complicated Hamiltonians, the method can only be implemented numerically. For instance, Montrunich et al.[19] studied the quantum phase transition in the two-dimensional random transverse-field Ising model. In analogy with the one-dimensional case,[22,23] they found an infinite randomness critical point, but the critical exponent take different values. Schehr and Rieger[136] studied the interplay between dissipation, quantum fluctuations and disorder in the random transverse-field Ising chain coupled to dissipative baths. In agreement with theoretical predictions,[30,31] they found that the dissipation freezes the quantum dynamics of large, locally ordered clusters which then dominate the low-energy behavior. This leads to a smearing of the quantum phase transition.[30]

Let us also mention a class of methods that are not numerically exact, but have greatly fostered our understanding of quantum many-particle systems: the dynamical mean-field theory (DMFT). Its development started with the pioneering work of Metzner and Vollhardt[137] on the Hubbard model in infinite dimensions. The basic idea of this approach is a natural generalization of the classical mean-field theories to quantum problems: The quantum many-particle Hamiltonian is reduced to a quantum impurity problem coupled to one or several self-consistent baths.[138] This impurity problem is then solved self-consistently, either by approximate analytical or by numerical methods. In contrast to classical mean-field theories such as Hartree-Fock, the DMFT contains the full local quantum dynamics. (This means that the DMFT suppresses spatial fluctuations but keeps the local imaginary-time fluctuations.) By now, DMFT methods have been applied to a wide variety of problems ranging from model Hamiltonians of strongly correlated electrons to complex materials. For instance, the DMFT was instrumental in understanding the Mott metal-insulator phase transition in the Hubbard model. More recently, the DMFT has been combined with realistic band structure calculations to investigate many-particle effects and strong correlations in real materials. Reviews of these developments can be found, e.g., in Refs. 139 and 140.

Let us finally point out that we have focused on bulk quantum phase transitions. Impurity quantum phase transitions[42] require a separate discussion that is mostly beyond the scope of this chapter (Note, however, that within the DMFT method a bulk quantum many-particle system is mapped onto a self-consistent quantum impurity model.) Some of the methods discussed here such as quantum Monte Carlo can be adapted to impurity problems. Moreover, there are powerful special methods dedicated to impurity problems, most notably Wilson's numerical renormalization group.[141,142,143]

# SUMMARY AND CONCLUSIONS

In this chapter, we have discussed quantum phase transitions. These are transitions that occur at zero temperature when a nonthermal external parameter such as pressure, magnetic field, or chemical composition is changed. They are driven by quantum fluctuations which are a consequence of Heisenberg's uncertainty principle. At first glance, it might appear that investigating such special points in the phase diagram at the absolute zero of temperature is purely of academic interest. However, in recent years, it has become clear that the presence of quantum phase transitions has profound consequences for the experimental behavior of many condensed matter systems. In fact, quantum phase transitions have emerged as a new ordering principle for low-energy phenomena that allows us to explore regions of the phase diagram where more conventional pictures based on small perturbations about simple reference states are not sufficient.

In the first part of the chapter, we have given a concise introduction to the theory of quantum phase transitions. We have contrasted the contributions of thermal and quantum fluctuations, and we have explained how their interplay leads to a very rich structure of the phase diagram in the vicinity of a quantum phase transition. It turns out that the Landau-Ginzburg-Wilson (LGW) approach that has formed that basis for most modern phase transition theories can be generalized to quantum phase transitions be including the imaginary time as an additional coordinate of the system. This leads to the idea of the quantum-to-classical mapping which relates a quantum phase transition in $d$ space dimensions to a classical one in $d+1$ dimensions. We have also briefly discussed situations in which the LGW order parameter approach can break down, a topic that has attracted considerable interest lately.

The second part of this chapter has been devoted to computational approaches to quantum phase transitions with the emphasis being on Monte-Carlo methods. If one is mainly interested in finding the universal critical behavior (i.e., the overall scaling scenario, the critical exponents, and the critical amplitude ratios), then often a purely classical simulation scheme based on quantum-to-classical mapping is most efficient. We have illustrated this approach for the transverse-field Ising model with and without dissipation, for the bilayer Heisenberg antiferromagnet, and for dirty bosons in two dimensions. If one is interested in nonuniversal questions such as quantitative results for critical coupling constants or observables, a true quantum algorithm must be used. We have reviewed several quantum Monte Carlo approaches to quantum spin and boson Hamiltonians and discussed their results for the quantum phase transitions in these systems. We have also considered fermionic systems and the extra complications brought about by the generic appearance of the notorious sign problem.

At present, it is probably fair to say that Monte Carlo simulations of model systems that are free of the sign problem (bosons, spin systems without frustration, and some special fermionic systems) have become so powerful that the properties of their quantum phase transitions can be determined quantitatively with high precision (see for instance the accuracy of some of the exponent values quoted in the preceding sections). In contrast, for many frustrated spin systems, the results are limited to a qualitative level, and for quantum phase transitions in generic fermionic systems (with sign problem), direct computational attacks are still of limited utility.

## ACKNOWLEDGEMENTS

Over the years, the author has greatly benefited from discussions with many friends and colleagues, in particular, D. Belitz, A. Castro-Neto, A. Chubukov, P. Coleman, K. Damle, V. Dobrosavljevic, P. Goldbart, M. Greven, S. Haas, J. A. Hoyos, F. Iglói, T. R. Kirkpatrick, A. Millis, D. Morr, M. Norman, P. Phillips, H. Rieger, S. Sachdev, A. Sandvik, J. Schmalian, Q. Si, R. Sknepnek, G. Steward, J. Toner, M. Vojta, A. P. Young.

This work has been supported in part by the NSF under grant nos. DMR-0339147 and PHY99-07949, by Research Corporation and by the University of Missouri Research Board. Parts of this work have been performed at the Aspen Center for Physics and the Kavli Institute for Theoretical Physics, Santa Barbara.

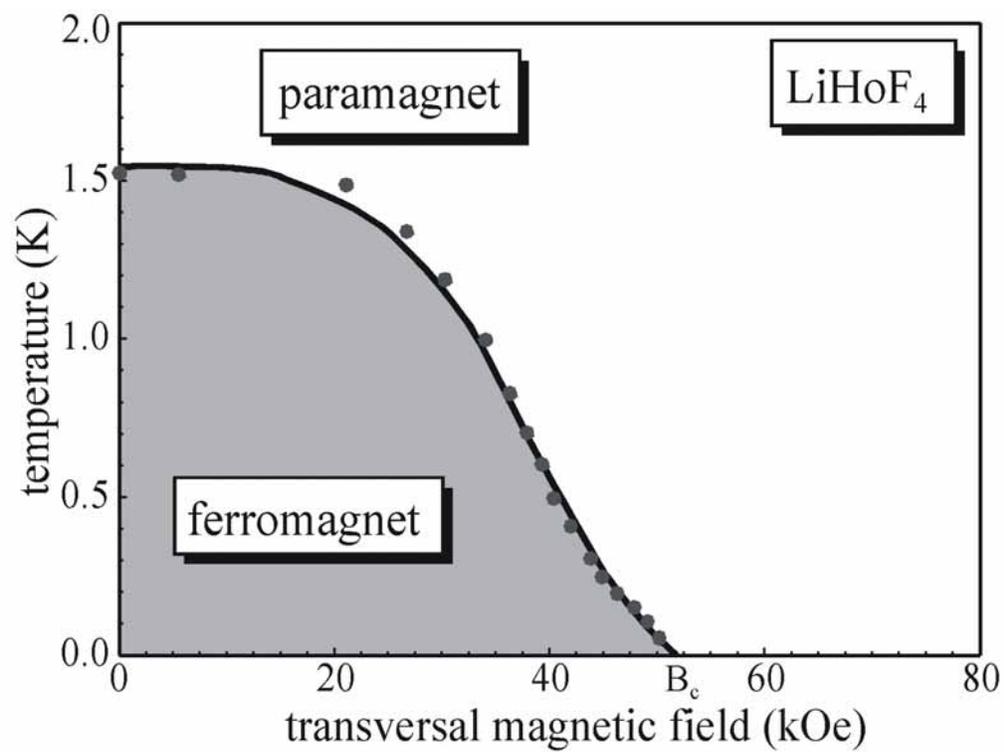

**Figure 1:** Phase diagram of LiHoF4 as function of temperature and transverse magnetic field.

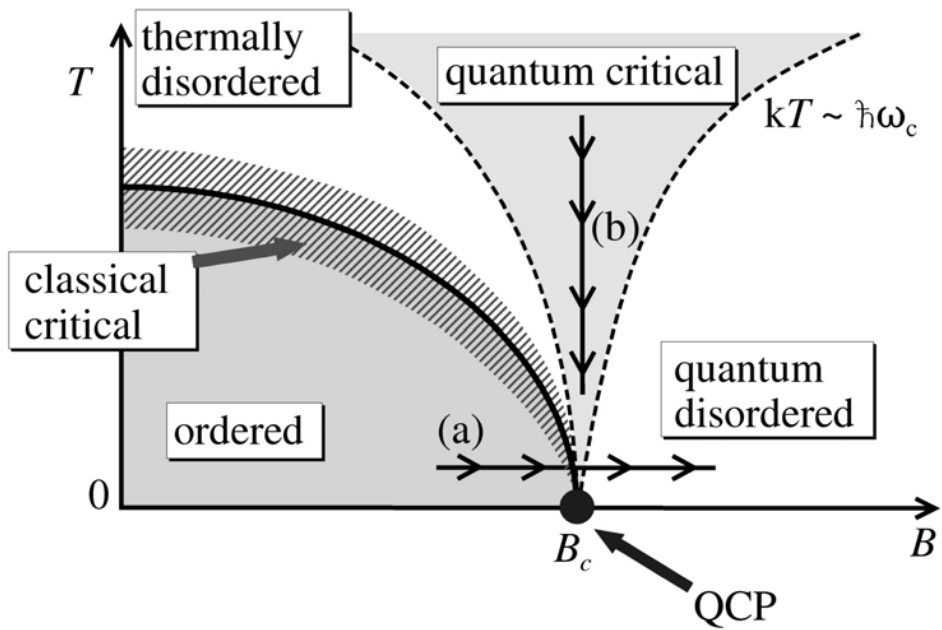

**Figure 2:** Schematic phase diagram close to a quantum critical point for systems that have an ordered phase at nonzero temperature. The solid line is the finite-temperature phase boundary while the dashed lines are crossover lines separating different regions within the disordered phase. QCP denotes the quantum critical point.

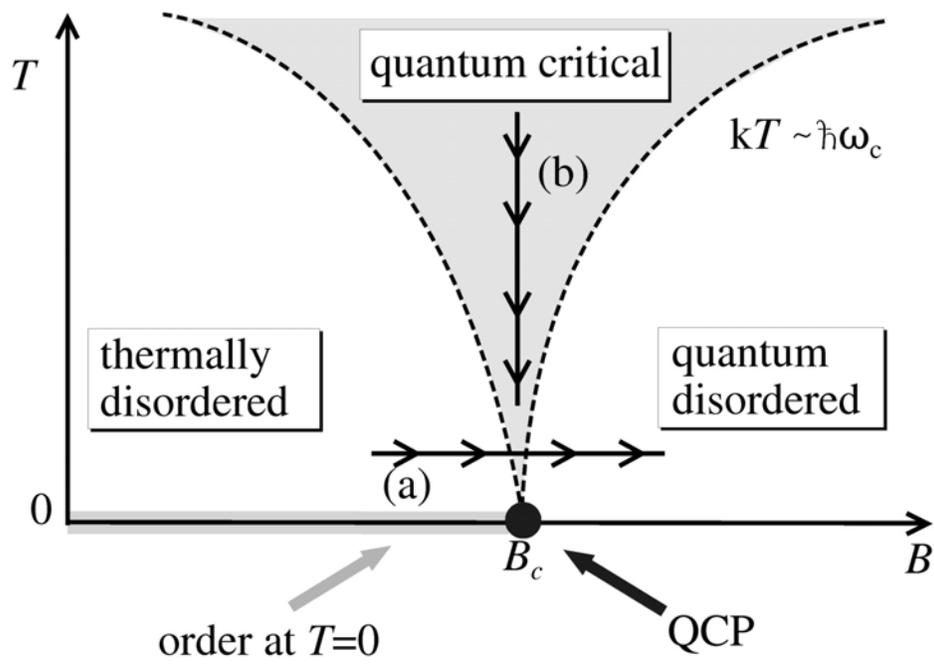

**Figure 3:** Same as Figure 2, but for systems that display long-range order only at exactly zero temperature.

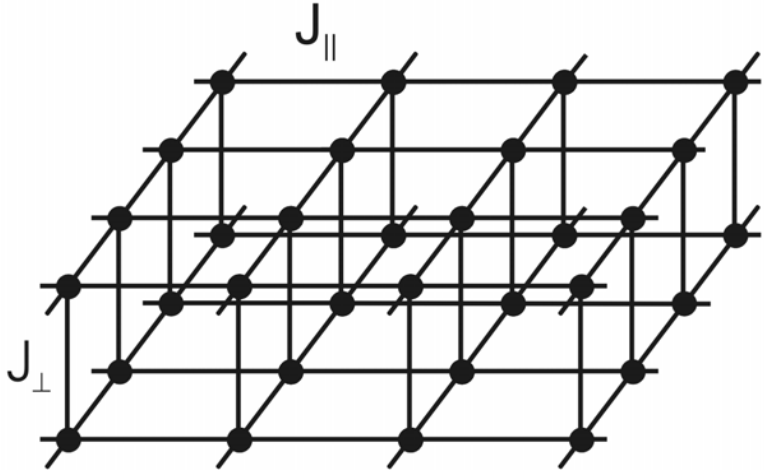

**Figure 4:** Sketch of the bilayer Heisenberg quantum antiferromagnet. Each lattice site is occupied by a quantum spin ½.

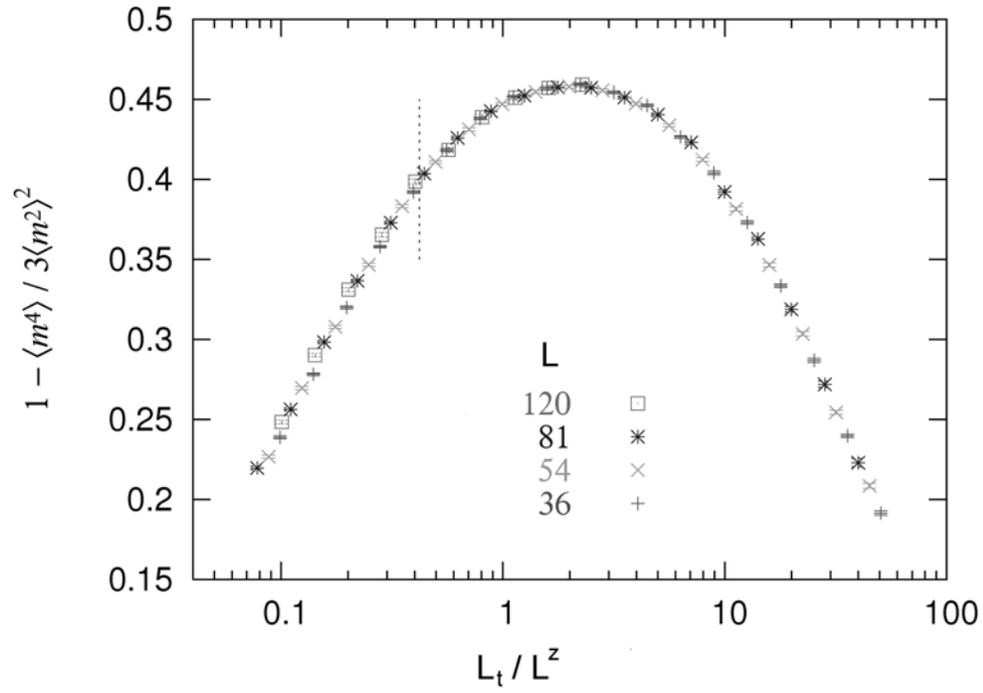

**Figure 5:** Scaling analysis of the Binder cumulant *B* of the classical Hamiltonian [24] at criticality ($\alpha = 0.6$, $\varepsilon J = 0.00111$, $K = 1.153$) with a dynamical exponent $z = 2$ (from Ref. 64).

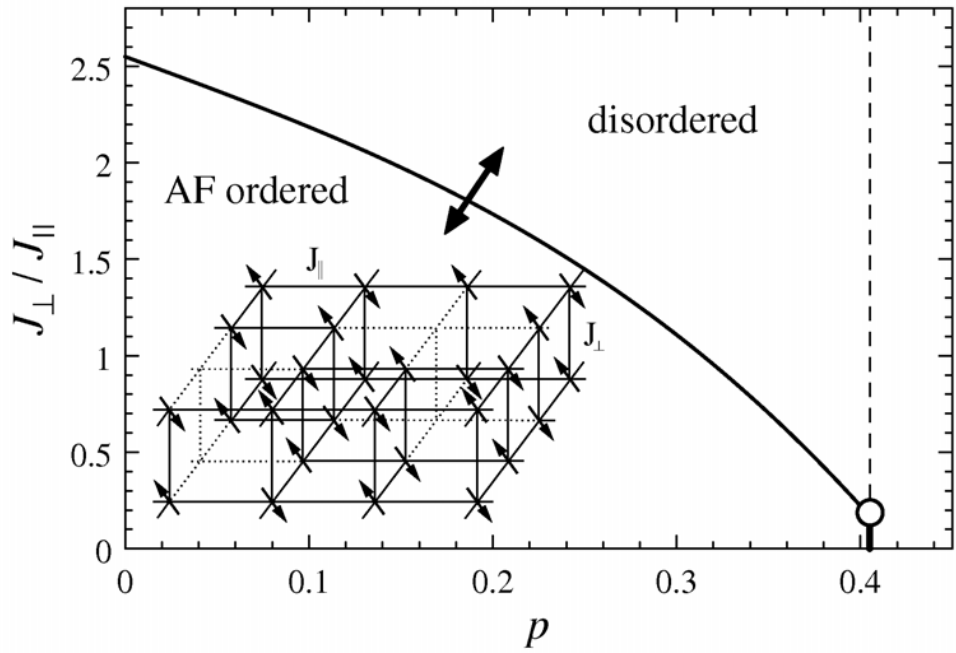

**Figure 6:** Phase diagram of the dimer-diluted bilayer Heisenberg antiferromagnet, as function of $J_\perp / J_\parallel$ and dilution $p$. The dashed line is the percolation threshold; the open dot is the multicritical point. (from Ref. 71).

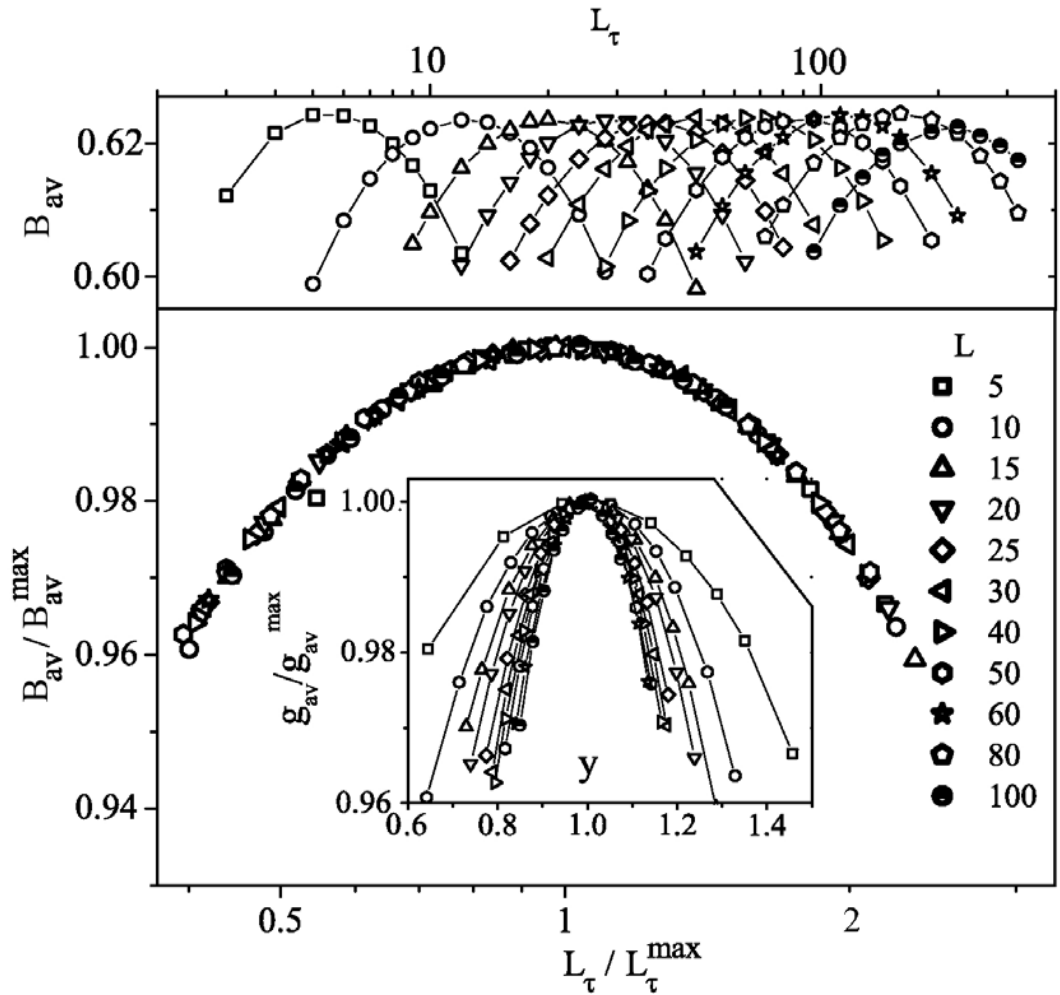

**Figure 7**: Upper panel: Binder cumulant of the classical Hamiltonian [29] at the critical point as a function of $L_t$ for various $L$ and impurity concentration $p= 1/5$. Lower panel: Power-law scaling plot of the Binder cumulant. Inset: Activated scaling plot of the Binder cumulant (from Ref. 71).

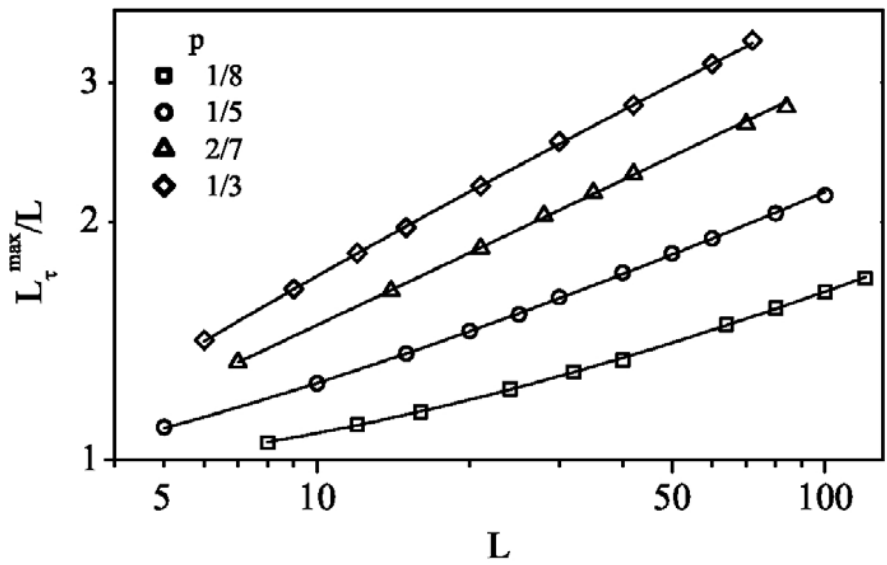

**Figure 8:** $L_t^{max}$ vs. $L$ for four impurity concentrations. The solid lines are fits to $L_t^{max} = aL^z(1+bL^{-\omega})$ with $z = 1.31$, $\omega = 0.48$ (from Ref. 71).

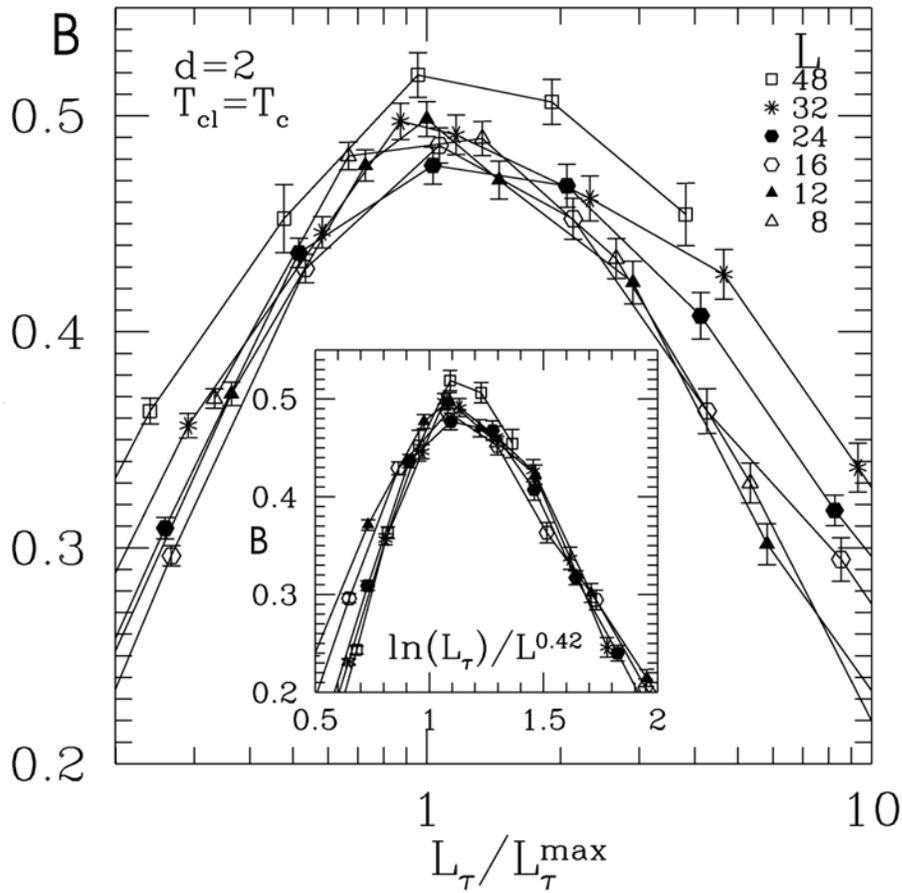

**Figure 9**: Binder cumulant of the classical Hamiltonian [31] at the critical point. Main panel: Power-law scaling plot. Inset: Scaling plot according to activated scaling (from Ref. 79).

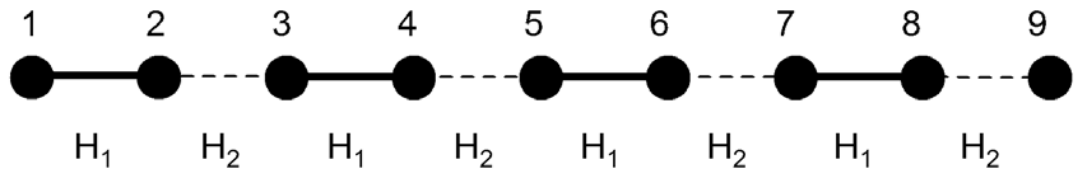

**Figure 10:** Checkerboard decomposition of the one-dimensional XXZ Hamiltonian.

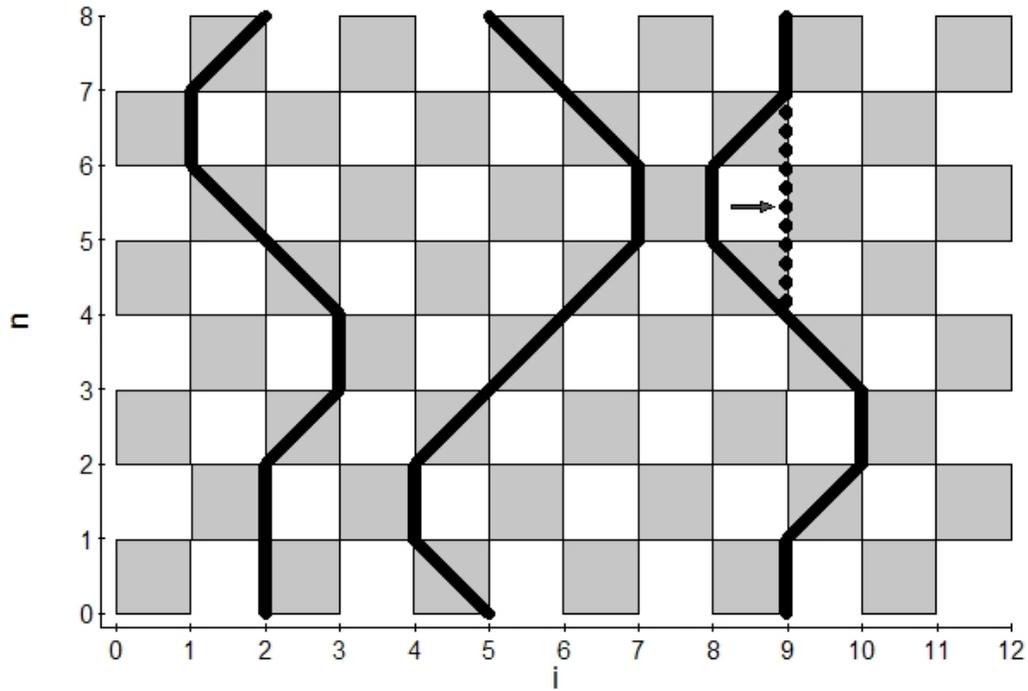

**Figure 11:** World-line configuration for the XXZ Hamiltonian [38]. The world lines (thick lines) connect space-time points where the *z*-component of the spin points up. They can either be straight or cross the shaded squares which show where the imaginary time evolution operators $e^{-\varepsilon \hat{H}_1}$ and $e^{-\varepsilon \hat{H}_2}$ act. The doted line shows the configuration change after a local Monte Carlo update.

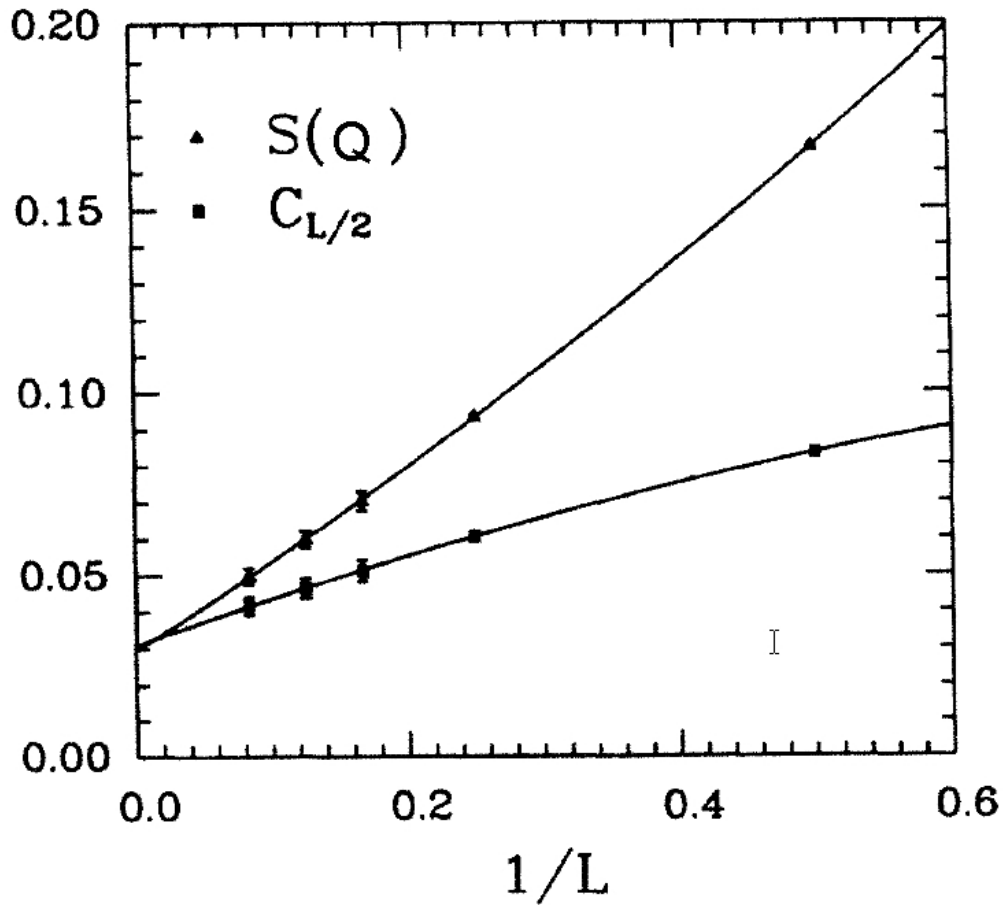

**Figure 12:** World-line Monte Carlo results for the square lattice Heisenberg antiferromagnet: Structure factor and the long-distance limit of the correlation function as functions of the linear system size $L$. The intercept on the vertical axis can be used to find the staggered magnetization (from Ref. 104).

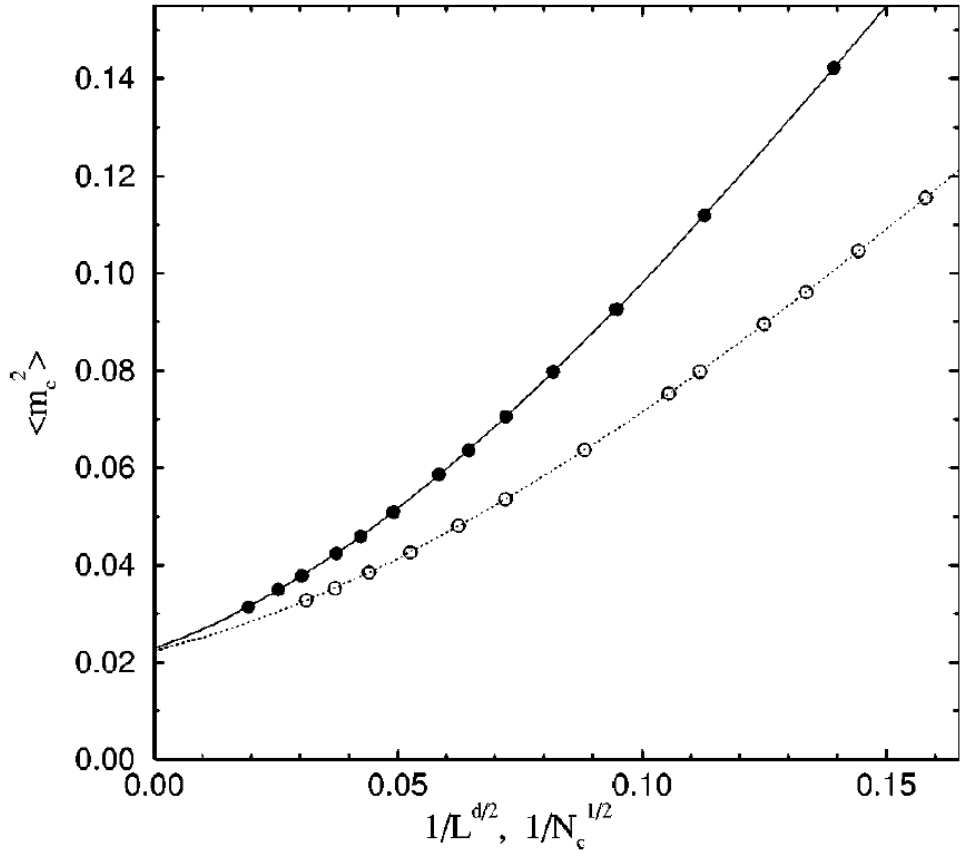

**Figure 13:** Squared staggered ground state magnetization of the Heisenberg model on a site-diluted lattice at $p=p_p$. The two curves correspond to two different ways of constructing the percolation clusters in the simulation. Solid circles: largest cluster on $L \times L$ lattices, open circles: clusters with a fixed number $N_c$ sites (from Ref. 107).

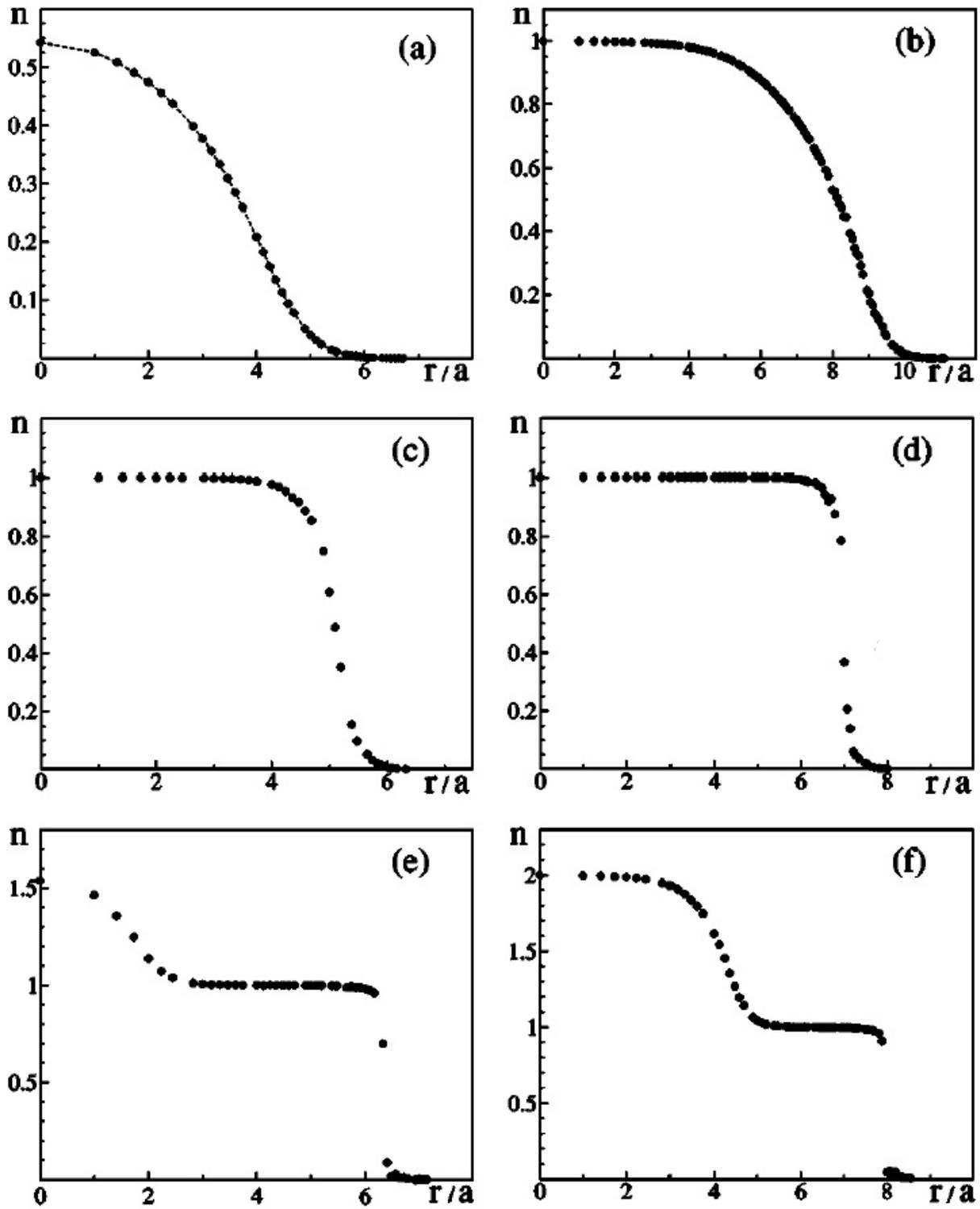

**Figure 14:** Superfluid-insulator transition in an optical lattice: Particle density (per lattice site) as function of the distance from the trap center for various parameters and filling factors (from Ref. 111).

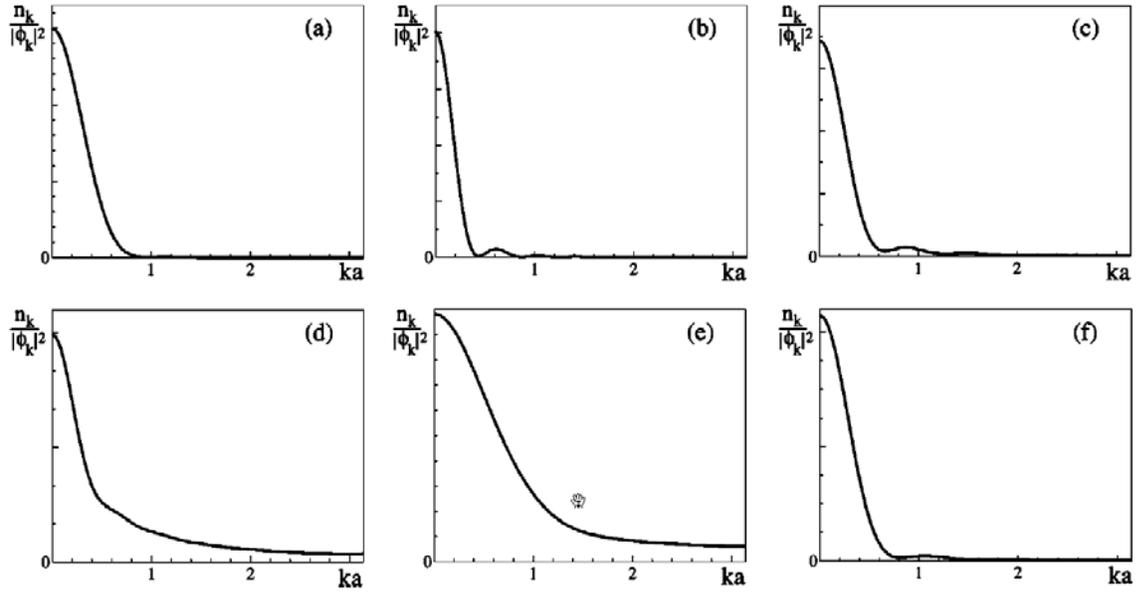

**Figure 15:** Superfluid-insulator transition in an optical lattice: Single-particle momentum distribution. Panels (a) to (f) correspond to the systems shown in Fig. 14 (from Ref. 111).